\documentclass[letterpaper,twocolumn,10pt]{article}

\usepackage{usenix-2020-09}
\usepackage[utf8]{inputenc}
\usepackage[T1]{fontenc}
\usepackage{xcolor}
\usepackage{filecontents}
\usepackage{times}  
\usepackage{helvet} 
\usepackage{courier} 
\usepackage{balance} 
\usepackage{url}
\usepackage{wrapfig}
\usepackage{graphicx}
\usepackage{textcomp}
\usepackage{subcaption}
\usepackage{comment}
\usepackage{tikz}
\usepackage{amsfonts}
\usepackage{mathrsfs}
\usepackage{makecell}
\usepackage{bm}
\usepackage{amsmath,amsfonts,amsthm,mathtools} 
\usepackage{amssymb}

\usepackage{bbm}
\usepackage[framemethod=TikZ]{mdframed}
\usepackage{hyperref}
\usepackage{booktabs}
\usepackage{multicol}
\usepackage{multirow}
\usepackage{colortbl}
\usepackage{tcolorbox}
\usepackage{soul}
\usepackage{wrapfig}
\usepackage{xspace}
\usepackage{bbding}
\usepackage{pifont}
\usepackage[numbers,sort]{natbib} 
\usepackage{wasysym}
\usepackage{threeparttable}
\usepackage[linesnumbered,ruled,vlined]{algorithm2e} 
\usepackage{cleveref}
\usepackage{makecell}
\usepackage{xcolor}
\usepackage{threeparttable}
\usepackage{acronym}
\usepackage{xltabular}

\usepackage{marvosym}
\usepackage[available]{usenixbadges}

\makeatletter
\renewcommand{\@fnsymbol}[1]{}
\makeatother

\definecolor{lightpink}{rgb}{1.0, 0.95, 0.95}
\definecolor{lightblue}{rgb}{0.87, 0.87, 1.0}

\newcommand{\tokenizer}{\cT}
\newcommand{\tokenstring}{\mathbf{t}}

\newcommand{\llm}{f}

\newcommand{\transformer}{\psi}

\newcommand{\firstlayer}[2]{\transformer^{#1}_{#2}}

\newcommand{\interstatesalone}{\bh}
\newcommand{\interstates}[2]{\bh^{#1}_{#2}}

\newcommand{\inpemb}[1]{\cE^{#1}}
\newcommand{\inpembmat}[1]{\bw_{in}^{#1}}
\newcommand{\inpembdim}{d_{in}}

\newcommand{\maxtokennum}{N_\tokenizer}

\newcommand{\numepochs}{E}

\newcommand{\inputtext}{x}
\newcommand{\advdata}{\cX_\cA}
\newcommand{\invmodel}{\theta}

\newcommand{\semanticeval}{S}
\newcommand{\distance}{\mathrm{d}}
\newcommand{\varchangetbs}{\varphi_\bz}

\acrodef{ai}[AI]{Artificial Intelligence}
\acrodef{pii}[PII]{Personally Identifiable Information}
\acrodef{nlp}[NLP]{Natural Language Processing}
\acrodef{llm}[LLM]{Large Language Model}
\acrodef{lm}[LM]{Language Model}
\acrodef{dp}[DP]{Differential Privacy}
\acrodef{tee}[TEE]{Trusted Execution Environment}

\acrodef{cossim}[CS]{Cosine Similarity}
\newcommand{\rouge}{ROUGE}

\newcommand{\llama}{Llama}

\newcommand{\qwen}{Qwen}
\newcommand{\norobots}{NoRobots}

\acrodef{mse}[MSE]{Mean Squared Error}
\acrodef{is}[IS]{Internal State}

\mdfdefinestyle{customstyle}{%
    backgroundcolor=white,
    linewidth=1pt,        
    frametitlefont=\bfseries,
    roundcorner=5pt,      
    innertopmargin=10pt,   
    innerbottommargin=10pt,   
    nobreak=true
}

\newcommand{\lesson}[1]{
\begin{mdframed}[style=customstyle] 
\textbf{Takeaway:}  
{#1}
\end{mdframed}
}

\newcommand{\eg}{\hbox{{e.g.}}\xspace}
\newcommand{\resp}{\hbox{{resp.}}\xspace}
\newcommand{\ie}{\hbox{{i.e.}}\xspace}

\newcommand{\aka}{\hbox{{a.k.a.}}\xspace}

\newcommand{\bheading}[1]{{\vspace{4pt}\noindent{\textbf{#1}}}}
\newcommand{\iheading}[1]{{\vspace{4pt}\noindent{\textit{#1}}}} 

\renewcommand{\epsilon}{\varepsilon}

\def\:#1{\protect \ifmmode {\mathbf{#1}} \else {\textbf{#1}} \fi}

\newcommand{\bh}{\mathbf{h}}

\newcommand{\bw}{\mathbf{w}}

\newcommand{\bz}{\mathbf{z}}

\newcommand{\bB}{\mathbf{B}}

\newcommand{\bU}{\mathbf{U}}
\newcommand{\bV}{\mathbf{V}}

\newcommand{\bZ}{\mathbf{Z}}

\newcommand{\cA}{\mathcal{A}}

\newcommand{\cE}{\mathcal{E}}

\newcommand{\cL}{\mathcal{L}}

\newcommand{\cP}{\mathcal{P}}

\newcommand{\cT}{\mathcal{T}}

\newcommand{\cV}{\mathcal{V}}
\newcommand{\cW}{\mathcal{W}}
\newcommand{\cX}{\mathcal{X}}

\newcommand{\cZ}{\mathcal{Z}}

\renewcommand{\epsilon}{\varepsilon}

\DeclareMathOperator*{\argmin}{arg\,min} 
\DeclareMathOperator*{\argmax}{arg\,max}

\DeclarePairedDelimiterX{\inp}[2]{\langle}{\rangle}{#1, #2}

\DeclareMathOperator*{\expectedvalue}{\mathbb{E}}

\newcommand{\norm}[1]{\left\lVert#1\right\rVert}
\newcommand{\abs}[1]{\left\lvert#1\right\rvert}

\newcommand{\Real}{\mathbb{R}}

\theoremstyle{definition}

\def\BibTeX{{\rm B\kern-.05em{\sc i\kern-.025em b}\kern-.08em
    T\kern-.1667em\lower.7ex\hbox{E}\kern-.125emX}}

\begin{document}

\title{Depth Gives a False Sense of Privacy: LLM Internal States Inversion}

\author{
{\rm Tian Dong\textsuperscript{1}, Yan Meng\textsuperscript{1,\Letter}\thanks{\textsuperscript{\Letter}Yan Meng and Haojin Zhu are corresponding authors.}, Shaofeng Li\textsuperscript{2}, Guoxing Chen\textsuperscript{1}, Zhen Liu\textsuperscript{1}, and Haojin Zhu\textsuperscript{1,\Letter}}\\
{\textsuperscript{1}Shanghai Jiao Tong University, \{tian.dong, yan\_meng, guoxingchen, zhu-hj\}@sjtu.edu.cn}\\
{\textsuperscript{2}Southeast University, shaofengli@seu.edu.cn}
}

\maketitle

\begin{abstract}
Large Language Models (LLMs) are increasingly integrated into daily routines, yet they raise significant privacy and safety concerns.
Recent research proposes collaborative inference, which outsources the early-layer inference to ensure data locality, and introduces model safety auditing based on inner neuron patterns.
Both techniques expose the LLM's \textit{Internal States} (ISs), which are traditionally considered irreversible to inputs due to optimization challenges and the highly abstract representations in deep layers.
In this work, we challenge this assumption by proposing four inversion attacks that significantly improve the semantic similarity and token matching rate of inverted inputs. Specifically, we first develop \textit{two white-box optimization-based attacks} tailored for low-depth and high-depth ISs. These attacks avoid local minima convergence, a limitation observed in prior work, through a two-phase inversion process.
Then, we extend our optimization attack under more practical black-box weight access by leveraging the transferability between the source and the derived LLMs.
Additionally, we introduce a \textit{generation-based attack} that treats inversion as a translation task, employing an inversion model to reconstruct inputs.
Extensive evaluation of short and long prompts from medical consulting and coding assistance datasets and 6 LLMs validates the effectiveness of our inversion attacks.
Notably, a 4,112-token long medical consulting prompt can be nearly perfectly inverted with 86.88 F1 token matching from the middle layer of \llama-3 model.
Finally, we evaluate four practical defenses that we found cannot perfectly prevent ISs inversion and draw conclusions for future mitigation design.

\end{abstract}

\section{Introduction}

Despite its widespread application, the large size of \acp{llm} prohibits fast inference on local devices, forcing users to send their inputs (\aka, prompts) to the cloud and risk privacy leakage.
This also impedes the application in sensitive domains and commercial cooperation~\cite{musk_ban_apple}. Moreover, as the model scale continues to grow (\eg, Llama-3 has a size up to 405B~\cite{dubey2024llama}), a single server can merely load the model in one piece, let alone swift inference.

Therefore, collaborative inference~\cite{lin2024splitlora,mei2024helix,zhang2024edgeshard} has been widely applied to enforce \textit{data locality}, where the shallow layers are stored on the local device and only the \acp{is} are transmitted to the cloud for continuous inference on rest layers.
Meanwhile, to meet the requirements of trustworthy \ac{ai}~\cite{white_house_order}, \acp{is} can also be exposed to a third party for safety auditing, as \acp{is} of deep layers can be leveraged to robustly identify factual errors~\cite{azaria-mitchell-2023-internal,chen2024inside,su-etal-2024-unsupervised,DBLP:journals/corr/abs-2406-15927,ye2024physics}, defend jailbreaks, backdoors~\cite{li2024safety,lamparth2024analyzing,dong2024trojaningplugins}, or manipulate internal representations of the model's concepts~\cite{zou2023representation,
burns2023discovering,chen2024states}.

The potential exposure of increasingly used \acp{is} raises our research question: \textit{Can we invert the input query based on the \acp{is}, even in highly deep \acp{llm}?}
Current embedding inversion~\cite{song2020information} assigns trainable variables to each input token and selects the candidate tokens via optimization, which is proven effective on conventional \acp{lm} (\eg, BERT).
Recent works show that text embeddings or model outputs~\cite{song2020information,morris-etal-2023-text,zhang2024extracting} can be used to invert inputs.
These attacks train generative inversion models conditioned on observed embeddings or outputs.

Yet, simple adoption cannot work well for \acp{is} because of two new challenges.
First, \acp{is} are designed for subsequent inference and contain abstract logical representations~\cite{chen2024states}, which are inherently different from previously studied embeddings or model outputs of high semantically relevance with inputs.
Second, \acp{llm} have significantly more layers, higher width, and larger dictionary than \acp{lm} studied in prior work, which further hinders the inversion, especially for \acp{is} of deep layers because of feature loss based on the information bottleneck~\cite{infobottleneck17}.
Therefore, we need more powerful inversion attacks to evaluate the privacy risk of \acp{is}.

In this work, we are the first to explore the inversion feasibility of \acp{is} by proposing both optimization-based and generation-based attacks adapting to white-box access and black-box access to model weights.
Specifically, our white-box attacks are designed for the adversary (\eg, curious-but-honest inference server) who can exploit the weights to optimize the input text with nearly exact and correctly ordered tokens without any assumption on input distribution.
Our black-box attacks are suitable for a third-party adversary (\eg, \ac{llm} auditor) who can probe the \acp{is} for analysis and can train an inversion model based on her own surrogate data of similar distribution to the victim's queries.

Since searching for the optimal token sequence through brute force is infeasible, we introduce a novel two-phase inversion for the optimization-based attack: we first invert the input embeddings and then recover the correct input tokens.
For shallow layers, our attack, Embedding Recovery (ER), produces embeddings of candidate inputs by minimizing the distance of its \acp{is} to the target.
Then, the tokens with the closest input embedding to the optimized embedding are selected.
This tackles the large-dictionary challenge by avoiding searching over significantly huge token combination space.
For deep layers, ER can fail because of gradient explosion.
We propose Token Basis Selection (TBS) that determines the optimal combination among base vectors of input embedding space as the inverted embeddings for further token inversion.
This tackles the high-depth challenge by reducing optimized variables and avoiding local minima encountered in the previous solution~\cite{song2020information}.

Without access to the target model weights, we first extend our optimization attacks to the black-box setting by identifying whether the target is derived from adversary-known \acp{llm}, based on our insights that a large number of \acp{llm} are derived from existing ones instead of pretrained from scratch.
For the generation-based attack, we regard the \acp{is} as an encoded language and use the encoder-decoder models, which are commonly used in machine translation, for input inversion.
To tackle the challenge of representation discrepancy between \acp{is} and semantic meaning, we propose a projection module that aligns the \acp{is} with the encoder for inversion with the decoder.

Our evaluations include 6 real-world high-ranking \acp{llm}, both short-context prompts, as adopted in existing works, and additional long-context prompts on medical consulting and coding assistance.
The results demonstrate the inversion effectiveness.
For example, given \acp{is} from the middle layer of \texttt{\llama-3-8B-Instruct}, our TBS attack can invert input of 4,112 tokens with 86.88 F1 token matching and 95.19 semantic similarity (see \Cref{fig:example-p1,fig:example-p2,fig:example-p3}) which cannot be reached by prior work.
Our generation-based attack can also achieve 81.6 F1 score for inputs of medium length (\ie, $\sim$1k tokens) which is comparable to the white-box attack.
Lastly, we test four defenses including quantization, dropout, noisy input embedding, and \ac{dp} through the Laplace mechanism.
Our black-box attack cannot be mitigated without greatly deteriorating the model utility, calling for more effective defenses in the future.

In summary, our contributions are:
\begin{itemize}
    \item We are the first to systematically investigate the input inversion risk of \ac{llm} \acp{is}. Our work reveals that an attacker can successfully recover sensitive prompts of LLMs, spanning up to 4,112 tokens, from their \acp{is}.
    
    \item To overcome the challenges of semantic spasticity and feature loss from high-depth layers, we propose four novel inversion attacks adapting to both white-box and black-box attack settings.
    
    \item We extensively evaluate our attacks on sensitive inputs including medical dialogues and coding assistance. We also evaluated \ac{dp}-based defense and found our attack can still invert input of high semantic similarity even significantly sacrificing the downstream inference quality.
\end{itemize}

\section{Preliminaries \& Motivation}
In this section, we first briefly introduce the modern \ac{llm} implementation, the risk of \ac{is} exposure and overview existing inversion techniques.

\subsection{Language Modeling}

We begin with a brief recap of how modern \acp{llm} processes the texts. Formally, an input text (\aka, prompt) $\inputtext$ is first tokenized by the tokenizer $\tokenizer_\llm$ of an \ac{llm} $\llm$ into a string of tokens $\tokenizer_\llm(x)\coloneq \tokenstring^x=[t_i^x]_{\{i=0,\cdots, \maxtokennum\}}$ where each token is marked with an ID $t_i$ bounded by the maximum token number $\maxtokennum$, \ie, $t_i\in[0,\cdots,\maxtokennum]$.
The token IDs are then mapped by the input embedding layer $\inpemb{\llm}$ of weight $\inpembmat{\llm}\in\Real^{\maxtokennum \times \inpembdim}$ into the sub-matrix of corresponding row vectors $\inpemb{\llm}(\tokenstring^x)\coloneq [\inpembmat{\llm}(t_i^x)]_{\{i=0,\cdots, N_t\}}$.
We denote the first $l$ Transformer layers of an \ac{llm} $\llm$ is $\firstlayer{\llm}{l}$.
The \acp{is} of the $l$-th layer of an \ac{llm} are $\interstates{\llm}{l}$.
Put together, given input text $x$, the \acp{is} at $l$-th layer of an \ac{llm} $\llm$ are $\interstates{\llm}{l}(x)=\firstlayer{\llm}{l}(\inpemb{\llm}(\tokenizer_\llm(x))$.

\begin{table*}[t]
\centering
\caption{Comparison with previous inversion attacks against (large) language models.}
\label{tab:comparison}
\resizebox{\linewidth}{!}{
\begin{tabular}{@{}ccccccccc@{}}
\toprule
\multirow{2}{*}{\textbf{Inversion Target}} & \multirow{2}{*}{\textbf{Method}} & \multirow{2}{*}{\textbf{\begin{tabular}[c]{@{}c@{}}Weight\\ Access\end{tabular}}} & \multicolumn{3}{c}{\textbf{Inversion Goal}} & \multicolumn{3}{c}{\textbf{Evaluation}} \\ \cmidrule(l){4-9} 
 &  &  & Semantic-preserving & Token Matching & Attribute Inference & Max. Model Size$^1$ & Data Types & Max. Data Length \\ \midrule

Embeddings~\cite{pan_sp20,song2020information,morris-etal-2023-text,li2023sentence} & \begin{tabular}[c]{@{}c@{}}Generation\\ Optimization\end{tabular} & \begin{tabular}[c]{@{}c@{}}\CIRCLE\\ \Circle\end{tabular} & \begin{tabular}[c]{@{}c@{}}\ding{51}\\ \ding{55}\end{tabular} & \begin{tabular}[c]{@{}c@{}}\ding{55}\\ \ding{51}\end{tabular} & \begin{tabular}[c]{@{}c@{}}\ding{55}\\ \ding{51}\end{tabular} & 110M & Chats, Medicine & 422 \\ \midrule
Outputs~\cite{morris2024language,zhang2024extracting} & Generation & \CIRCLE & \ding{51} & \ding{55} & \ding{55} & 7B & System Prompts, Chats & 256 \\ \midrule
Gradients$^2$~\cite{DBLP:conf/emnlp/DengWLWSLRD21,feng2024uncovering, ccs24unveiling} & Optimization & \Circle & \ding{51} & \ding{51} & \ding{55} & 7B & Chats, Code, Math & 512 \\ \midrule
Internal States (Ours) & \begin{tabular}[c]{@{}c@{}}Generation\\ Optimization\end{tabular} & \begin{tabular}[c]{@{}c@{}}\CIRCLE\\ \Circle\end{tabular} & \begin{tabular}[c]{@{}c@{}}\ding{51}\\ \ding{55}\end{tabular} & \begin{tabular}[c]{@{}c@{}}\ding{55}\\ \ding{51}\end{tabular} & \ding{55} & 70B & Medicine, Code & 4,112 \\ \bottomrule
\end{tabular}
}
\begin{tablenotes}
\footnotesize
\item[1] $^1$ Only for models with publicly known size. $^2$ Attacks targeting training data instead of input texts.
\end{tablenotes}
\end{table*}

\subsection{Motivation}

In \ac{llm} service, maintaining the confidentiality of \acp{is} is not always guaranteed. We identify two scenarios in which \acp{is} may be exposed to an untrusted party for inversion.

\textbf{\ac{llm} Alignment \& Concept Engineering.}
\acp{llm} are notorious for hallucination, which can deceive and misinform the user. Therefore, \acp{llm} are persistently surveilled for safety reasons. Recently, a growing number of studies~\cite{chen2024inside,su-etal-2024-unsupervised,li2024safety,DBLP:journals/corr/abs-2406-15927,chen2024states,zhou-etal-2024-alignment} show that the \acp{is} are robust indicators of hallucination. This provides \ac{llm} holders with a promising solution to correct model behavior in the runtime~\cite{ye2024physics}.
For instance, instead of directly examining the prompts, OpenAI may run an automated safety classifier to improve their services based on the classifier-generated metadata based on policy~\cite{openai_business_datause}.
Besides, \acp{is} are also required in representation engineering~\cite{zou2023representation} to control the model concept.

\textbf{Collaborative \& TEE-shielded LLM Inference.}
Several solutions~\cite{borzunov-etal-2023-petals,lin2024splitlora,mei2024helix,zhang2024edgeshard} have been proposed for layer-wise \ac{llm} splitting and partitioning to accelerate \ac{llm} serving capacity.
For instance, 
EdgeShard~\cite{zhang2024edgeshard} dynamically shards models onto edge devices for closer \ac{llm} deployment to the data source.
PETAL~\cite{borzunov-etal-2023-petals} allows several servers to collaboratively infer or finetune models up to 405B through layer-wise model splitting.
HELIX~\cite{mei2024helix} exploits heterogeneous GPUs from different regions in the globe.
Besides, the split model can be loaded in \ac{tee} to protect input privacy~\cite{sun2023sp,Zhang:2024:odml}.
In both settings, the party holding rest model layers receives and infers \acp{is}.

\subsection{Challenges}
There has been a line of work studying how the \acp{lm} leak the user input texts.
For example, the text embeddings are shown at risk of leaking the input text through inversion attacks in both white-box and black-box settings~\cite{morris-etal-2023-text,song2020information}.
Recently, it has been shown that the user's input prompt can be accurately reconstructed with only the \ac{llm} logits~\cite{morris2024language} or outputs~\cite{zhang2024extracting}.
The gradients in federated learning~\cite{DBLP:conf/emnlp/DengWLWSLRD21,feng2024uncovering, ccs24unveiling} can also leak the training texts by inversion attacks.
Technically, the inversion either attempts to locate the candidate input tokens (without considering their relative positions) through optimization (Optimization-based), or trains a generative model with surrogate data for inversion conditioned on observed embeddings (Generation-based).

Our work is the first to explore the feasibility of input inversion through \acp{is} as an \ac{llm} inner representation.
Applying existing embedding inversion techniques on \acp{is} can result in poor inversion due to the difference between text embeddings and \ac{llm} \acp{is}.
We identify two new challenges:

\textit{Increased model scale and token dictionary size.}
In general, \acp{llm} have a deeper layered architecture, higher model width, and larger token dictionary than conventional pretrained \acp{lm}, making it almost impossible to get the exact input as the victim through the dictionary attack.
For example, typical base BERT only contains 110M parameters, 12 layers with width 768 and around 30k dictionary size, while modern 8B \llama-3 contains 32 layers of width 4,096 and more than 120k dictionary tokens.
The larger model scale significantly increases trainable parameters, causing previous optimization-based inversion~\cite{song2020information} to fail because of falling into local minima on highly compressed \acp{is} (see \Cref{sec:attackmethod}).

\textit{Inference-oriented and complex representations.}
Previous embedding inversion attacks focused on sentence embeddings which are typically optimized for semantic relevance based on mean-pooled encoder outputs~\cite{ni2021large}.
On the contrary, in the context of \acp{llm}, the \acp{is} are generated for continuous inference, thus contain more sparse semantic features than semantically-enhanced embeddings.
In addition, the \acp{is} of deep  layers contain a higher level of concept abstraction~\cite{zou2023representation,concept2025coling} like reasoning, which further increases the difficulty of accurate input inversion.

In this work, we systematically analyze the inversion risk of \acp{is} by addressing the aforementioned challenges through novel optimization-based and generation-based attacks. More importantly, we validate the effectiveness of these attacks in practical settings, as summarized in \Cref{tab:comparison}.

\section{Threat Model}
\label{sec:threatmodel}

We consider a curious-but-honest adversary.
For example, a malicious third-party auditor authorized to access \acp{is} for safety auditing~\cite{azaria-mitchell-2023-internal,chen2024inside} can stealthily store the observed \acp{is} for the offline inversion.
In the context of collaborative inference, the adversary server hosting middle layers or bottom layers of the deployed \ac{llm}, receives \acp{is} of the splitting layer and sends results to the next party following the protocol with the client and other servers~\cite{borzunov-etal-2023-petals,mei2024helix}.
The server can allocate partial computational power to invert the user queries based on knowledge of splitting layers and observed \acp{is}.

\begin{figure}
    \centering
    \includegraphics[width=0.9\linewidth]{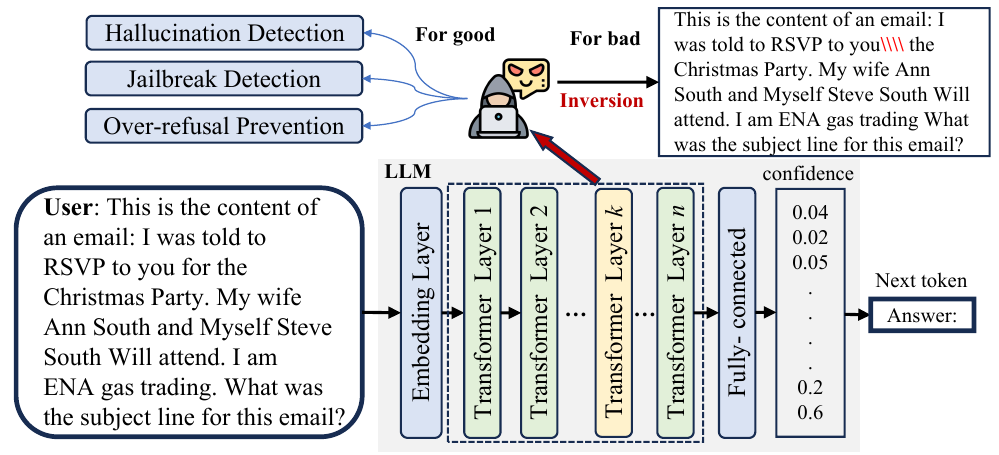}
    \caption{A curious-but-honest \ac{llm} safety auditor or collaborative inference party can observe \acp{is} and recover the nearly \textit{exact} user inputs even in deep layers (false inverted tokens are in red).}
    \label{fig:threat}
\end{figure}

\bheading{Adversary Goals.}
The adversary $\cA$ aims to invert the input texts/prompts of victim $\cV$ based on the observed \acp{is} $\interstates{\llm_\cV}{l}(\inputtext)$ of ground truth inputs $\inputtext$ and her model $\llm_\cA$.
The inverted texts should preserve the semantics and exact tokens as the victim's input.
Note that this goal is harder than privacy attribute inference and can be applied for further analysis (\eg, \ac{pii} or user identification~\cite{huang2024authorship}).
Formally, the \acp{is} inversion is
\begin{equation}
\begin{aligned}
    & \hat{\inputtext}=\argmin_{\inputtext'\in \text{dom} \tokenizer} \distance(\interstates{\llm_\cA}{l}(\inputtext'), \interstates{\llm_\cV}{l}(\inputtext)), \\ & \mathrm{ s.t. }  
\semanticeval(\inputtext', \inputtext) \geq \tau_\semanticeval, \abs{\tokenizer_{\llm_\cA}(\inputtext') \cap \tokenizer_{\llm_\cA}(\inputtext)} / \abs{\tokenizer_{\llm_\cA}(\inputtext)} \geq \tau_{tm}
\end{aligned}
\end{equation}
where $\text{dom} \tokenizer$ denotes the tokenizer's input domain, $\distance$ is the distance metric between two \acp{is}, $\semanticeval$ evaluates semantic similarity, $\tau_\semanticeval$ (\resp, $\tau_{tm}$) is a threshold of $\semanticeval$ (\resp, token matching).

\bheading{Adversary Knowledge.}
We assume the adversary knows the layer $l$ of \acp{is} and consider two settings:
1) The adversary has white-box access to the weights (\ie, $\llm_\cA=\llm_\cV$).
For instance, the collaborative inference PETAL~\cite{borzunov-etal-2023-petals} requires the adversary server to know the whole weights so as to select its layers during the load balancing~\cite{mei2024helix}, thus know the layer index for attacks;
2) The adversary has no knowledge (black-box access) of the model weights and can only observe the \acp{is}.
A typical example can be third-party model behavior auditing~\cite{azaria-mitchell-2023-internal,chen2024inside,ji2024llm,su-etal-2024-unsupervised} by probing \acp{is} of specific layers\footnote{\url{https://github.com/microsoft/TaskTracker/blob/main/task_tracker/utils/activations.py}} for the layer-specific detector training~\cite{taskdrift_satml25}.

\bheading{Adversary Capacities.}
We consider passive attacks thus the adversary cannot interact with user or manipulate the deployed model.
However, the adversary can store the observed \acp{is} $\interstates{}{l}$ (we omit $\llm_\cV$ and $\inputtext$ for simplicity) from the target model $\llm_\cV$ and have computational resources enough for inversion but insufficient for brute force search.
The adversary can also query the target model to obtain \acp{is} of her owned data $\advdata$ which is of distribution similar to but not exactly as the victim's query.
For example, the adversary may have general instruction tuning data but the victim queries are from specific domains (\eg, coding), which are not known by the adversary because of no interaction.

\section{Internal States Inversion}
\label{sec:attackmethod}
In this section, we first overview the attacks and clarify the terminology.
Then, we elaborate our attack insight and method.

\subsection{Overview}

\begin{figure}[t]
    \centering
    \includegraphics[width=0.95\linewidth]{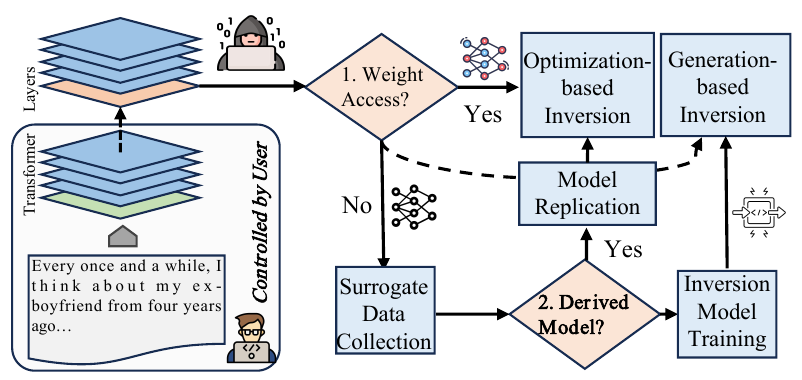}
    \caption{Overview procedure of our attacks. Depending on the model access, the adversary can adopt our optimization-based or generation-based attacks to achieve attack goals.}
    \label{fig:attack-overview}
\end{figure}

Our attack framework proceeds according to the adversary's knowledge to the target \ac{llm}.
\Cref{fig:attack-overview} shows the overall workflow of our attacks based on the adversary's capacities.
In particular, in cases of white-box access to the model weights, our \textit{optimization-based attack} iteratively updates the inverted input by matching the observed target \acp{is}.
One important advantage is that, there is no assumption on the knowledge of victim's prompt domain and length (\ie, \textit{data-agnostic}).
We propose two attacks, Embedding Recovery (ER) and Token Basis Selection (TBS), targeting shallow and deep \acp{is} respectively.

Without access to the weights, the adversary follows our black-box attacks.
Due to high cost of pretraining, the target \ac{llm} is likely to be derived from public open-source models through finetuning or merging.
As a result, the first step is to determine whether the target model is derived from any known base \acp{llm} type.
If it is, the adversary trains a surrogate model and apply our optimization-based attacks.
If not, our \textit{generation-based} attack trains an inversion model based on the \acp{is} queried by the adversary's surrogate data.
To tackle the challenge of semantic irrelevance, we introduce a projection module based on sparse encoder~\cite{zhang2024extracting} and translation model to enhance the inversion.
This can be weaker than previous white-box attacks because the knowledge of query data can limit the inversion accuracy in case of distinct data distribution.
Note that the black-box generation-based attack can also work on open-sourced \acp{llm}, as long as the adversary only needs to observe ISs to train inversion models.

\subsection{White-box Inversion}
We first introduce the strawman approach, then present our proposed two white-box attacks, Embedding Recovery (ER) and Token Basis Selection (TBS), for inverting \acp{is} of shallow layers and deep layers, respectively.

\begin{figure}
    \centering
    \includegraphics[width=0.99\linewidth]{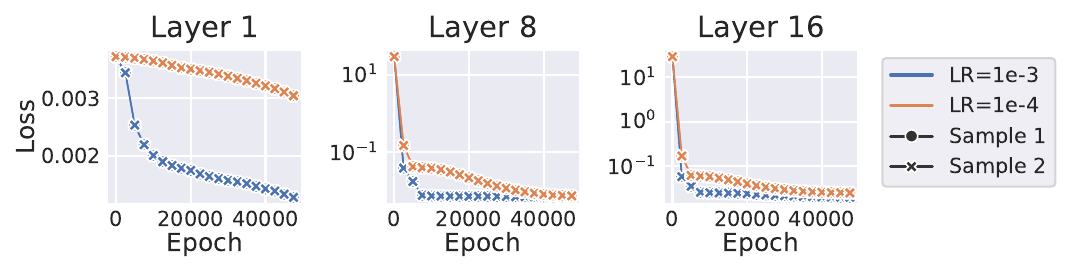}
    \caption{Evaluation of strawman attack TS on \texttt{\llama-3-8B-Instruct}.
    The strawman approach, TS, fails to converge on the deeper layers even under improved settings. Moreover, when attacking the first layer, the inverted texts contain no overlapping tokens with the input texts.}
    \label{fig:ts-loss}
\end{figure}
\bheading{Strawman Approach.}
Inspired by previous optimization-based embedding inversion~\cite{song2020information,dpforward_ccs23}, the strawman attack, Token Selection (TS), typically assigns trainable variables $\bZ=[\bz_1,\cdots, \bz_{\abs{\interstates{}{l}}}]\in\Real^{\abs{\interstates{}{l}}\times \maxtokennum}$ of each embeddings to invert $i$-th input token of by $\argmax \widehat{\bz}_i$, where $\widehat{\bz}_i$ are the rows of
\begin{equation}
   \widehat{\bZ}=\argmax_{\bZ}\norm{ \firstlayer{}{l}(\text{softmax}(\bZ  / T)\cdot \inpembmat{\llm_\cA}) - \interstates{\llm_\cV}{l}}_2,
\end{equation}
which is obtained through gradient-based optimization.
Nevertheless, TS can fail to locate candidate tokens for \acp{llm} with higher depth and larger token dictionary due to convergence into local minimum.
In \Cref{fig:ts-loss}, we apply TS on \acp{is} from different layers of \texttt{\llama-3-8B-Instruct} and evaluate different settings.
We observe that TS gets halted for deeper layers and output non-readable inverted texts (more quantitative evaluation in \Cref{subsec:evaluation-white-box}).

\begin{algorithm}[t]
\small
    \caption{ER (\colorbox{lightblue}{blue}) and TBS (\colorbox{lightpink}{red}) attacks}
    \label{alg:white-box}
    \KwIn{The adversary's model $\llm_\cA$ with input embedding weight $\inpembmat{\llm_\cA}$ and the first $l$ Transformer layers $\firstlayer{\llm_\cA}{l}$, the tokenizer $\tokenizer_{\llm_\cA}$, target hidden states $\interstates{\llm_\cA}{l}$ of length $\abs{\interstates{\llm_\cA}{l}}$, learning rate $\mu$, optimizer \texttt{optim}, distance metric \texttt{dist},  steps $E$}
    \KwOut{Inverted prompt $\hat{x}$.}
    Initialize $\cL\leftarrow[], \cZ\leftarrow[]$;

    \colorbox{lightblue}{
    Initialize $\widehat{\bw}\leftarrow\mathbf{0}$ where $\widehat{\bw}\in\Real^{\abs{\interstates{\llm_\cV}{l}}\times \inpembdim}$;
    }

    \colorbox{lightpink}{\parbox{0.9\linewidth}{
    $\bU, \Delta, \bV^\top \leftarrow \text{SVD}(\inpembmat{\llm_\cA})$ where $\bV\in\Real^{\inpembdim\times \inpembdim}$.
    Set $\bB\leftarrow\cV^\top$;

    \If{Apply Unbiased Basis}{
        $\bB\leftarrow \cV$;
    }

    Initialize projection weights $\bz\leftarrow [\frac{1}{\inpembdim}]$ and $\bz\in\Real^{\abs{\interstates{\llm_\cV}{l}}\times \inpembdim}$;
    }}

    \tcp{1. Optimization.}

    \For{$i \gets 1$ \KwTo $\numepochs$}{

        \colorbox{lightpink}{
        $\widehat{\bw}\leftarrow\varchangetbs(\bz\cdot \bB)$;
        }
        
        Compute $\cL_{inv}$ with \Cref{eq:inversion_total} and save as $\cL[i]$.

        \colorbox{lightblue}{$\widehat{\bw}\leftarrow\texttt{optim}(\cL[i], \widehat{\bw}, \mu)$;
        $\cZ[i]\leftarrow \widehat{\bw}$;}
        \colorbox{lightpink}{
        $\bz\leftarrow\texttt{optim}(\cL[i], \bz, \mu)$;
        $\cZ[i]\leftarrow \bz$;}

    }

    \tcp{2. Prompt Inversion.}

    \colorbox{lightblue}{
        $\widehat{\bw}\leftarrow\cZ[\argmin_i{\cL}]$;
    }
    
    \colorbox{lightpink}{
        $\widehat{\bw}\leftarrow\varchangetbs(\cZ[\argmin_i{\cL}]\cdot \bB)$;
    }
    
    $\widehat{\bw}\leftarrow \cW[\argmin_i{\cL}]$\; 
    $\tokenstring\leftarrow \argmax_{row} [\widehat{\bw} \cdot (\inpembmat{\llm_\cA})^\top / ( \norm{\widehat{\bw}}_{RN} \odot \norm{\inpembmat{\llm_\cA}}_{RN})] $;

    \Return $\tokenizer_{\llm_\cA}.\texttt{decode}(\tokenstring)$;

\end{algorithm}

\bheading{Our Approach.}
Instead of direct token inversion, our intuition is to first approximate the dummy input embeddings $\widehat{\bw}$ that matches with the adversary-observed \acp{is} $\interstates{\llm_\cV}{l}$, and then invert the candidate tokens as those of highest cosine similarity with $\widehat{\bw}$.
\Cref{alg:white-box} shows the overview of ER and TBS attacks, where the blue (\resp, red) blocks refer to the ER (\resp, TBS) attack, and the rest is shared by two attacks.
Note that the model weights are fixed during optimization.

\iheading{Embedding Recovery (ER).}
As shown in \Cref{fig:gradient_norm} of \Cref{subsec:evaluation-white-box}, for shallow layers, the gradients on dummy embedding $\nabla\widehat{\bw}$ are of smaller magnitude, thus can avoid local minimum with stable convergence.
Therefore, we directly optimize $\widehat{\bw}$ for \acp{is} matching with $\cL_{im} \coloneqq  \distance(\firstlayer{\llm_\cA}{l}(\widehat{\bw}), \interstates{\llm_\cV}{l})$.
Common choices of $\distance$ include $\cL_p$ norm and cosine distance.

In experiments, we notice overfitting during inversion, which results in the optimized $\widehat{\bw}$ of higher norm than embeddings from $\inpembmat{\llm_\cA}$ and causes incorrect token recovery in the second phase.
Thus, to ensure $\widehat{\bw}$ is similarly distributed to $\inpembmat{\llm_\cA}$, we introduce a penalty term based on distribution matching~\cite{dc_dm2023}:
$\cL_{dm} =  \expectedvalue_{\varphi\sim \cP_\varphi}\norm{\overline{\varphi(\widehat{\bw})} - \overline{\varphi(\inpembmat{\llm_\cA})}}_2$, where $\varphi$ denotes random Gaussian neural network as universal feature extractor and the overline denotes averaging. 
In practice, we sample random embedding batches of equal size from $\widehat{\bw}$ and $\inpembmat{\llm_\cV}$ before averaging in each step.
In total, the inversion loss is:
\begin{equation}
\label{eq:inversion_total}
    \cL_{inv} = \cL_{im} + \lambda\cdot \cL_{dm},
\end{equation}
where $\lambda$ balances the inversion and the penalty.

\begin{figure}[t]
    \centering
    \includegraphics[width=0.8\linewidth]{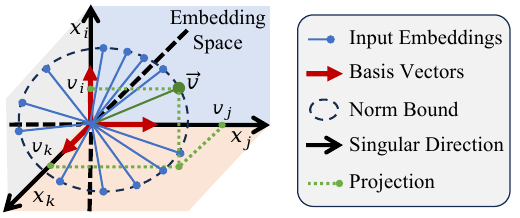}
    \caption{Intuition behind our TBS inversion attack. Instead of directly selecting candidate tokens from a large dictionary, our TBS attack optimizes weights among much fewer singular basis vectors (\eg, $v_i$, $v_j$ and $v_k$) to restore the candidate input embeddings (\eg, $\vec{v}$).}
    \label{fig:tbs-intuition}
\end{figure}

\begin{figure}[t]
    \centering
    \includegraphics[width=0.9\linewidth]{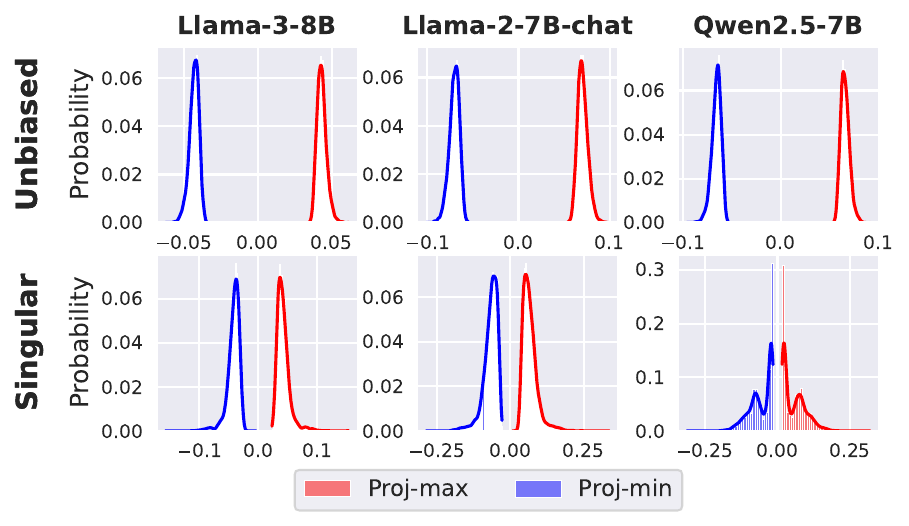}
    \caption{Comparison between unbiased basis and singular basis of \ac{llm}'s input embedding matrix for TBS attack. We show the histogram of maximum and minimum projection values of input embedding on the basis and found the singular basis is more biased towards particular token groups because of higher projected values.}
    \label{fig:motiv-unbiased}
\end{figure}

\iheading{Token Basis Selection (TBS).}
On deep layers, the gradients on the dummy embedding $\nabla\widehat{\bw}$ have increased magnitude because of more back-propagated gradients from previous layers accumulated by the chain rule, which destabilizes the inversion.
Therefore, our idea is to find the correct projection values $\widehat{\bz}$ of an orthogonal basis of $\inpembmat{\llm_\cA}$ to compose $\widehat{\bw}$, as illustrated in \Cref{fig:tbs-intuition}.
In contrast to TS, our TBS has much smaller search space.
For example, on \texttt{\llama-3-8B-Instruct}, the search space is reduced by 30 times (\ie, 4k for TBS v.s. 120k+ for TS to recover one token).
Compared with ER, the gradients on $\widehat{\bz}$ is stabilized because of an additional gradient term from the chain rule $\partial \widehat{\bw} / \partial \widehat{\bz}$ which is close to the small-scaled basis.
Inspired by TS, we also introduce a variable change function $\varchangetbs$ that bound the scale of $\bz$.
For example, in our experiments, we use arctan for $\varchangetbs$ to bound the $\bz\cdot\bB$.
More detailed analysis can be found in \Cref{app:additional_results}.

As the orthogonal basis is not unique, we propose two strategies: use the basis $\bV$ from SVD decomposition of $\inpembmat{\llm_\cA}$ or the $\bV^\top$ as an unbiased basis.
\Cref{fig:motiv-unbiased} compares the two strategies where we observe input embeddings are more centered around certain basis vectors of $\bV$ (\ie, higher projection values) where the projections of $\inpembmat{\llm_\cA}$ on unbiased basis $\bV^\top$ are more uniformly distributed (\ie, maximum projection values are bounded around 0.1).
We provide visualization to support the claim in \Cref{fig:visualization_emb_basis} of \Cref{app:additional_results}, and further compare two basis choices in \Cref{subsec:evaluation-white-box}.

\subsection{Black-box Inversion}
\label{sec:black-box}
The target \ac{llm} should be either finetuned or merged from an open-sourced base model or pretrained as a close-sourced model.
The adversary first identifies the model type based on the observed \acp{is} and then proceeds the extended optimization-based or generation-based inversion.

\bheading{Model Type Identification.}
Our insight is that the \acp{is} of derived models remain proximate to their base models because of minor weight perturbations during finetuning or merging.
As evidenced in \Cref{fig:tsne-llm}, \acp{is} exhibit tight intra-model clustering and clear inter-model separation across pretrained architectures.
To realize fine-grained detection, we design an ensemble autoencoder to check if the deployed model is derived from adversary-known open-sourced ones.
Specifically, the adversary queries a pretrained \ac{llm} $\llm$ with $\advdata$ and obtains corresponding \acp{is} $\interstates{\llm}{l}(\advdata)$.
The adversary trains an autoencoder on $\interstates{\llm}{l}(\advdata)$ to detect this \ac{llm} type.
By repeating this process on various pretrained models, the adversary trains a set of autoencoders $\{\phi_i\}_i$ for detecting \acp{llm} $\llm_i$.
In the test time, the adversary queries $\llm_\cV$ with $\cX_\cA$ to obtain given the target \acp{is} $\interstates{\llm_\cV}{l}(\cX_\cA)$ and harnesses the autoencoders $\{\phi_i\}_i$ to predict in an ensemble way:

\begin{equation}
    y\coloneqq\argmin_{i}\{\phi_i(\interstates{\llm_\cV}{l}(\cX_\cA))| \phi_i(\interstates{\llm_\cV}{l}(\cX_\cA)) \leq \tau\}.
\end{equation}

There are two possible outcomes.
First, the target \ac{llm} is derived from one of mainstream \acp{llm} via finetuning, adapter or model merging.
The adversary adopts this \ac{llm} for further attack. 
Second, there is no such \ac{llm}, and the model parameter is close-sourced.
Then, the adversary can reuse $(\advdata, \interstatesalone(\advdata))$ to train a generative inversion model.

\bheading{Extended Optimization-based Attack.}
The adversary aims to replicate the target model with a surrogate \ac{llm} $\llm_\cA$ that satisfies $\interstates{\llm_\cA}{l}(\inputtext)\approx \interstates{\llm_\cV}{l}(\inputtext)$ for any input $\inputtext$.
Inspired by the model distillation, we propose to replicate the \acp{is} with the pretrained base model $\llm_{base}$ of the identified type:
\begin{equation}
    \llm_\cA \coloneqq \arg \min _{\llm_{base}}\norm{\interstates{\llm_{base}}{l}(x_i)-\interstates{\llm_\cV}{l}(x_i)}_2, \forall x_i \in \advdata.
\end{equation}
Then, the adversary then applies our white-box attacks (\eg, TBS) onto the replicated model $\llm_\cA$ to invert input tokens of observed \acp{is} $\interstates{\llm_\cV}{l}$.

\bheading{Generation-based Attack.}
In case where the model is drastically different from the known \ac{llm} types, such as the target \ac{llm} is close-sourced, we propose to train an inversion model that translates the observed \acp{is} $\interstates{\llm_\cV}{l}$ into inputs.
\Cref{sec:threatmodel} assumes the adversary data $\advdata=\{\inputtext_i\}_i$ are of similar distribution to the victim's query, thus, before the attack, the adversary first trains an inversion model $\invmodel$ with previously obtained $(\cX_\cA, \interstates{\llm_\cV}{l}(\cX_\cA))$.

We use an encoder-decoder model $\theta$ for inversion because of its wide usage for translation and that the inversion essentially translates the observed \acp{is} into input tokens.
However, the \acp{is} are generated for continuous inference and deviate from the semantic meaning.
Besides, the target \acp{is} may have incompatible representation space with the encoder (\eg, different dimension).
Therefore, we propose to use a projection module $\varphi_{\invmodel}:\Real^{\inpembdim}\rightarrow \Real^{d_{enc}}$ on top of the inversion model to project the \acp{is} into the encoder's embedding space (width $d_{enc}$).
In total, the inverted input is
$\hat{\inputtext}\sim\invmodel(\varphi_{\invmodel}(\interstates{\llm_\cV}{l}))$.
Note that the projection module is necessary for aligning \acp{is}.
For instance, on the middle \acp{is} of GPT-2, the projection module brings 32.81\% inversion improvement (F1 score).

\bheading{Analysis from Bayesian Perspective.}
To better understand our generation-based inversion, we analyze the inversion based on Bayes' theorem.
The inversion can be formulated as $\argmax_\inputtext p(\inputtext | \interstatesalone, \invmodel)$, where we omit the projection $\varphi_\theta$ for simplicity.
We can apply the Bayes's theorem and obtain
\begin{equation}
    p(\inputtext | \interstatesalone, \invmodel) = \frac{p(\interstatesalone | \inputtext, \invmodel)  p(\inputtext | \invmodel) }{p(\interstatesalone |\invmodel)}.
\end{equation}
Notice that the observed \acp{is} $\interstatesalone$ and $\invmodel$ are known, thus the inversion can be formulated as:
\begin{equation}
     \argmax_\inputtext p(\interstatesalone | \inputtext, \invmodel)  p(\inputtext | \invmodel),
\end{equation}
where $p(\interstatesalone | \inputtext, \invmodel)$ means the posterior probability of $\interstatesalone$ given prompt input $\inputtext$ thus depends on the deployed \ac{llm}.
Therefore, the key is to maximize $p(\inputtext | \invmodel)$ where $\invmodel$ is dependent on the adversary's training data $\{(\inputtext_i, \interstatesalone_i)\}_i$.
Consequently, the adversary should minimize the distribution discrepancy between the adversary's data $\advdata$ and the victim's prompt input.

\section{Evaluation}
\label{sec:evaluation}

In this section, we evaluate the effectiveness of our attack and compare with previous approaches.
Then we investigate various attack settings.
Finally, we study potential mitigation.

\subsection{Experimental Setup.}

\bheading{Models.}
Throughout the evaluation, we mainly use the \texttt{\llama-3-8B-Instruct} (\llama-3)~\cite{dubey2024llama}, \texttt{\qwen2.5-7B-Instruct} (\qwen2.5)~\cite{hui2024qwen2} and \texttt{\qwen2.5-Coder-7B-Instruct} (\qwen2.5-Coder) as the most popular open-source \acp{llm}.
In addition, we also use \texttt{\llama-2-7B-Chat} (\llama-2) to fairly compare with prior work.
To evaluate our black-box attacks, we adopt \texttt{Bio-Medical-\llama-3-8B}~\cite{ContactDoctorBioMedicalLlama38B} (\llama-3-Doctor), which is a highly-downloaded model finetuned from \texttt{\llama-3-8B-Instruct} with medical instruction following data and \texttt{\llama-3.1-8B-Instruct} (\llama-3.1), which enhances \llama-3 in terms of multilingual capacities and context length via post-training~\cite{llama31}.
For larger models, we use \texttt{Llama-3.3-70B-Instruct} (Llama-3.3-70B).
The detailed statistics of base models are provided in \Cref{tab:stat_llm}.

\begin{table}[t]
\caption{The statistics of \ac{llm} considered in this work.}
\label{tab:stat_llm}
\resizebox{\linewidth}{!}{
\begin{tabular}{ccccc}
\toprule
\textbf{LLM} & \textbf{\# Params (B)} & \textbf{\# Token} & \textbf{Width} & \textbf{Depth} \\ \midrule
LLama-2-7B-chat & 6.74 & 32,000 & 4096 & 32 \\ \hline
Llama-3-8B-Instruct & 8.03 & 128,000 & 4096 & 32 \\ \hline
Llama-3.1-8B-Instruct & 8.03 & 128,000 & 4096 & 32
\\ \hline
\llama-3.3-70B-Instruct & 70.55 & 128,000 & 8192 & 80 \\ \hline
Qwen2.5-7B-Instruct & 7.62 & 151,643 & 3584 & 28 \\ \hline
Qwen2.5-Coder-7B-Instruct & 7.62 & 151,646 & 3584 & 28 
\\ \bottomrule
\end{tabular}}
\end{table}

\bheading{Metrics.}
We compute \ac{cossim} between the embeddings of original and inverted sentences.
To ensure reproducibility, we use an open-source embedding model \texttt{bge-large-en-v1.5}~\cite{bge_embedding} of edge-cutting performance on Massive Text Embedding Benchmark~\cite{muennighoff2022mteb,mteb_leaderboard}.
In addition, we use Exact Matching (EM) rate among the test dataset to evaluate the precision of our attacks.
We also use F1 score to evaluate the matched tokens by \llama-3's tokenizer.
Besides, we also consider widely used metrics BLEU and ROUGE to evaluate the semantic similarity.

\subsection{White-box Inversion Attack}
\label{subsec:evaluation-white-box}

In this section, we first evaluate our white-box attack on the Instruction-2M test data and compare with prior work.
Then, we show that our attacks can nearly fully invert the long prompt through case studies on long-context queries of medical consult and coding assistance.

\bheading{Attack Settings.}
We consider the layers of index 2, 4, 8, 16, 24 for the 32-layered \llama-3 models and consider the layer of index 14 for the 28-layered \qwen2.5 models.
We use the optimizer AdamW~\cite{loshchilov2018decoupled} with learning rate $\mu\in\{10^{-3}, 5\times 10^{-4}, 1\times 10^{-4}\}$.
We set the penalty weight $\lambda\in\{0, 10^{-3}\}$.
In default, we use the \ac{mse} to measure the distance between \acp{is} and the unbiased basis for our TBS attack.
We use scipy~\cite{2020SciPy-NMeth} to calculate the singular vectors in our TBS attack.
In our attacks, we initialize the dummy input embeddings with zero for ER and with $1/\inpembdim$ for the baseline TS and our TBS attacks.
We also consider two distance metrics between \acp{is} in the loss: \ac{mse} and cosine similarity (COS).
For simplicity, the default setting uses MSE as the distance metric, sets $\mu=5\times10^{-4}$ and $\lambda=0$, and $\numepochs=5\times10^4$.
For TBS attack, we use unbiased basis in default and set $\varchangetbs=\alpha\arctan(\cdot)$ with $\alpha=5/\pi$ to accelerate convergence.
In later sections, we will see that the optimal setting may vary according to the input data and the deployed model.

\begin{table}[t]
\centering
\caption{Comparison with previous attacks based on the \acp{is}, logits and outputs. We report the mean for each metric and the standard error of the mean (SEM).}
\label{tab:compare}
\resizebox{\linewidth}{!}{
\begin{tabular}{cccccc}
\toprule
\textbf{Attack} & \textbf{CS} & \textbf{BLEU} & \textbf{\rouge}& \textbf{EM} & \textbf{F1} \\ \hline
TS~\cite{song2020information} (L=2) & 44.24$\pm$ 0.32 & 0.00$\pm$ 0.00 & 0.00 & 0.00$\pm$ 0.00 & 0.00 $\pm$ 0.00 \\ \hline
logit2text~\cite{morris2024language} & 88.12$\pm$ 0.57 & 56.74$\pm$ 1.44 & 0.72 & 25.40$\pm$ 0.02 & 75.22 $\pm$ 1.00  \\ \hline
output2text~\cite{zhang2024extracting} & 93.26$\pm$ 0.33 & 55.00$\pm$ 1.46 & 0.77 & 16.21$\pm$ 0.02 & 75.94 $\pm$ 0.99   \\ \midrule
ER (L=2) & 94.24$\pm$ 0.47 & 74.89$\pm$ 1.56 & 0.87 & 52.61$\pm$ 0.02 & 88.22 $\pm$ 0.90  \\
ER (L=8) & 95.74$\pm$ 0.38 & 72.91$\pm$ 0.94 & 0.89 & 6.05$\pm$ 0.01 & 87.86 $\pm$ 0.69 \\ \midrule TBS (L=8) & 90.96$\pm$ 0.82 & 77.38$\pm$ 1.55 & 0.83 & 48.23$\pm$ 0.02 & 82.61 $\pm$ 1.44 \\ 
TBS (L=16) & 83.70$\pm$ 0.99 & 59.89$\pm$ 1.88 & 0.70 & 31.73$\pm$ 0.02 & 69.13 $\pm$ 1.72
\\TBS (L=24) & 88.07$\pm$ 0.83 & 65.47$\pm$ 1.69 & 0.77 & 27.35$\pm$ 0.02 & 75.32 $\pm$ 1.53
\\TBS (L=32) & 81.34$\pm$ 0.60 & 32.67$\pm$ 1.19 & 0.62 & 3.97$\pm$ 0.01 & 55.97 $\pm$ 1.02    \\\bottomrule
\end{tabular}
}
\end{table}

\subsubsection{Comparison}
We first compare our white-box attacks with previous attacks and later compare with our 
\Cref{tab:compare} shows the results.

\bheading{Baselines.}
We compare with output2text~\cite{zhang2024extracting} and 
logit2text~\cite{morris2024language} as two recent state-of-the-art inversion attacks against \acp{llm}, because the adversary (\eg, collaborative inference server) can also exploit the output and logits for inversion.
For fair comparison, we use their test dataset of Instruction-2M and manually verify 479 common samples from the two official implementations~\cite{morris2024language,zhang2024extracting}.
The maximum length of test samples is 63.
Besides, we also compare with TS baseline~\cite{song2020information}.

In \Cref{tab:compare}, we compare our proposed attacks with previous black-box and white-box attacks on different layer depths $L=2, 8, 16, 24$ on a 32-layered \llama-2.
Our ER attack can achieve high similarity scores on low layers and our TBS attack can remain effective even in deep layers like $L=24$.
We set $\mu=0.001$ for layer index lower than 8 and $\mu=0.01$ for the deeper layer to accelerate the convergence. 
Notably, our attacks achieve significant higher EM rates than prior work, indicating that the \acp{is} can reveal (almost) identical text information as the raw input text.

\iheading{Cause Analysis.}
We then proceed detailed analysis of inversion results.
As discussed in \Cref{sec:attackmethod}, TS directly selects candidate tokens for inverted input text, which does not work as the similarity scores between inverted and real input texts are close to zero.
In fact, the inversion loss converges around $10^{-2}$ and cannot be improved by tuning learning rates.
On the other hand, previous black-box attacks, exploiting the output logits or texts, result in comparable inversion in terms of semantics and token matching, and slightly lower EM score, because of the randomness during inversion model generation and higher information loss in the model output.

As for our attacks, ER performs better at shallow layers and TBS can remain effective in deep or even the close-to-last layers.
Notably, the ER attack can exactly invert more than half input texts on the shallow layer $L=2$.
This can be realistic in split learning with resource-constraint edge devices or small-sized enclave (\eg, SGX-v1) where the victim can only infer one or two shallow layers.

\begin{figure}[t]
    \centering
    \includegraphics[width=0.95\linewidth]{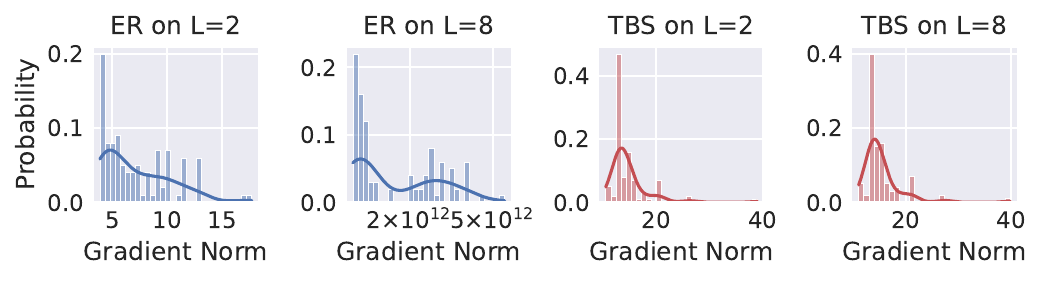}
    \caption{Distribution of gradient norm $\norm{\widehat{\bw}}_2$ and $\norm{\widehat{\bz}}_2$ for ER and TBS attacks, respectively.}
    \label{fig:gradient_norm}
\end{figure}

As for deep layers, the \acp{is} contain less input text features and are more difficult to invert.
Thus, even though the semantics are preserved, our ER attack's on deeper layer $L=8$ has significantly lower EM rate because of more noisy tokens (\eg, inversion as ``två017"'' for ``2/2/2017'') added during inversion.
From the optimization perspective, higher depth can lead to gradient exploding causing the directly recovered input embedding dissimilar to the actual token.

On the other hand, our TBS attack can stabilize the optimization curve by preserving the gradient magnitude.
In \Cref{fig:gradient_norm}, we testify this through distribution of the gradient norm on the test texts, where we can observe that ER on deep layer (L=8) generates gradient of magnitude $10^{12}$ while our TBS on deep layers can maintain the gradient norm of the similar magnitude as on shallow layers.
Therefore, the exploded gradients cause the convergence on local minimum and lead to low inversion quality.
Note that the gradients of TBS on L=8 have sightly larger norm than L=2, which explains why inversion on deeper layer should use lower learning rates.
As for qualitative results on the middle layer (L=16), on a random subset from Instruction-2M of 200 samples, ER only achieves CS score
50.35 and F1 score 7.56, which is significantly lower than our TBS attack shown in \Cref{tab:compare}.

\iheading{Results on Middle Layer.}
We note that the inversion by our TBS attack on the middle layer, although evaluated better in terms of EM rates, are worse than the deeper layer (L=24) in terms of semantic similarity scores and F1 scores.
We investigate the inverted texts and found that there are also numerous noisy tokens from other languages (\eg, Russian and Korean) replaced for the original English words.
More noisy tokens cause the readability degradation of inverted texts, thus lower the semantic similarity scores and token F1 score.
Although we tried lower learning rate $\mu=0.001$, the inversion on the middle layer does not get improved.
We suspect the reason lies in the optimization dynamics: the TBS inversion on L=16 gets saturated after 20,000 steps but keeps improving until steps 30,000 for L=24.
In future work, we will investigate real causes from the perspective of training dynamics.

\iheading{Inversion on Last Layer.}
We apply TBS attack on the last layer L=32.
As \Cref{tab:compare} shows, the depth further degrades inverted texts, which partially aligns with prior work~\cite{concept2025coling} that deep layers capture more complicated concepts instead of simple features of input texts.
Nonetheless, our attack validates a counterintuitive finding: high-depth latent layer can still leak the original inputs in entirety (EM$\approx$4).

\lesson{On short-context inputs, our optimization attacks outperform TS, achieve comparable inversion as previous generative inversion.}

\begin{figure*}[t]
    \centering
    \includegraphics[width=0.95\linewidth]{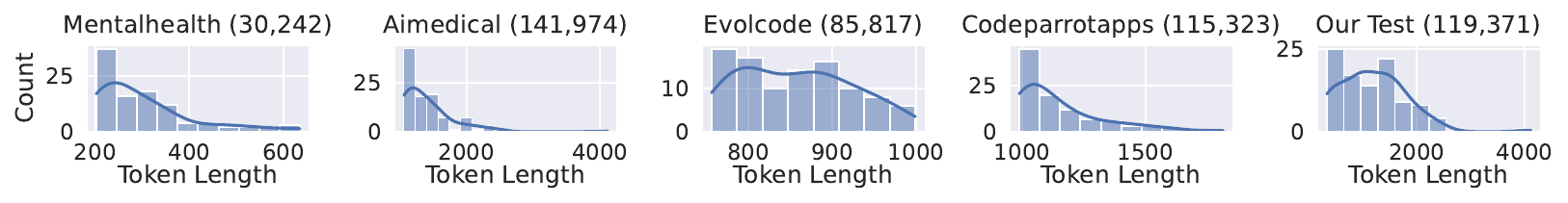}
    \caption{Distribution of token length from our test data (top 100 longest inputs of long-context benchmarks). The rightmost corresponds to the evaluation set for our TBS attack which contains longest 25 prompts from each benchmarks. We show the total token amount along in the figure title.}
    \label{fig:dataset-length}
\end{figure*}
\subsubsection{Case Studies: Long-context Inversion}

In the remainder of the paper, we evaluate how our inversion attack perform on long-context prompt through two case studies of privacy-sensitive tasks: healthcare consulting and coding assistance.
We use \llama-3 for its support of longer context and choose the middle layer (\ie, L=16) as it is reported best to probe the \ac{llm}~\cite{azaria-mitchell-2023-internal,burns2023discovering} and balances the inference cost between two parts in collaborative inference.
Thus, we mainly investigate our TBS attack.
In addition, we investigate optimal optimization strategies (\eg, distance metric in loss) for different tasks.

\bheading{Datasets.}
We consider datasets of two privacy-sensitive tasks: healthcare dialogue and coding assistance.
For medical data, we use the symptom descriptions from Aimedical~\cite{ai_medical_dataset} and Mentalhealth (MH) ~\cite{amod_2024}.
For coding assistance, we use two coding problem datasets Evolcode (EC) ~\cite{luo2023wizardcoder}
and CodeparrotApps (CA)~\cite{hendrycksapps2021} that contain prompts asking \ac{llm} to solve coding problem.
\Cref{fig:dataset-length} plots the distribution of length of top 100 long prompts for the four benchmarks, from which we can find that Aimedical contains the longest prompt of 4,112 tokens and CodeparrotApps has longer coding prompts than Evolcode.
We use the top 25 longest prompts from each dataset to evaluate our TBS (right-most of \Cref{fig:dataset-length}).

\begin{figure*}
    \centering
    \includegraphics[width=0.95\linewidth]{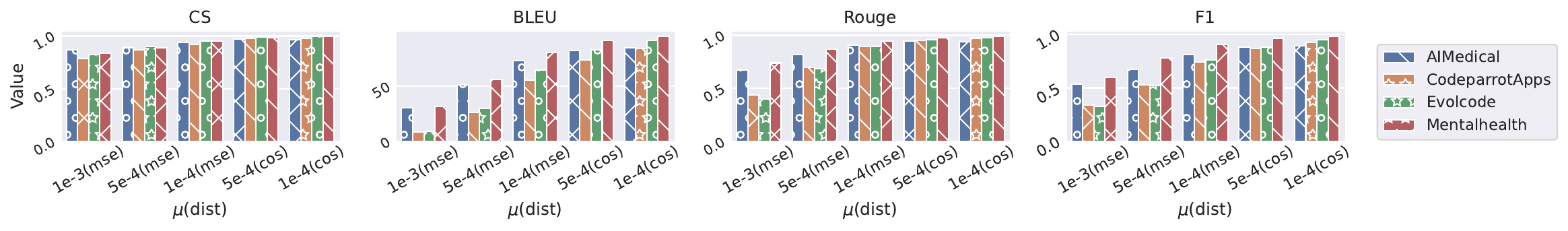}
    \caption{Evaluation of our TBS attack with different learning rates and distance functions (MSE and COS).}
    \label{fig:wb-long-lr-loss-dataset}
\end{figure*}

\bheading{Our attack can scale to long-context inversion.}
\Cref{fig:wb-long-lr-loss-dataset} compares the similarity metrics for different attack settings.
Note that due to long length and noisy inverted tokens, the EM rates are all zero, thus we omit the EM results to save space.
Specifically, the optimal settings are reported for lower learning rate $\mu=0.0001$ with COS as the distance metric, which leads to 99.4 CS and 97.88 F1 on Mentalhealth, and 98.12 CS and 96.7 F1 on Aimedical.
Both result in 0.99 Rouge score.
Next, we show examples of long-context inversion.

\begin{figure*}
    \centering
    \includegraphics[width=0.95\linewidth]{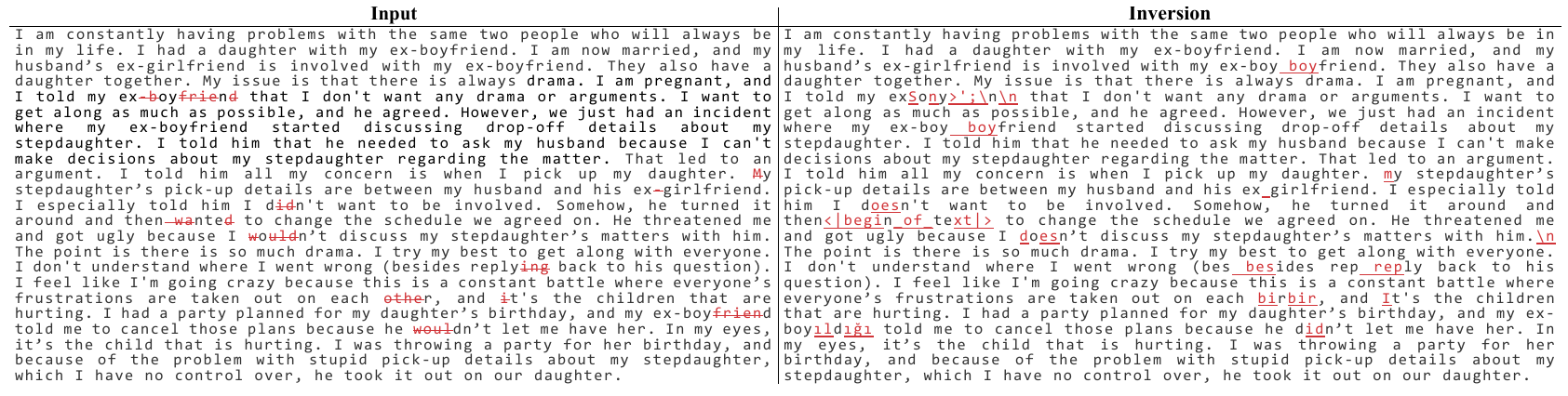}
    \caption{An inversion example from Mentalhealth consisting of 384 tokens. The missed tokens are highlighted in color.}
    \label{fig:example_health}
\end{figure*}

\bheading{Example: Healthcare Dialogue.}
We begin with examples of healthcare dialogues.
\Cref{fig:example_health} shows an inversion example of 384 tokens sampled from Mentalhealth.
The missing tokens are highlighted in color.
This example shows that our inversion attack can nearly invert the whole prompt text except for a small proportion of tokens.
Due to the space limit, we provide an example of inversion of 4,112 tokens in \Cref{fig:example-p1,fig:example-p2,fig:example-p3}.

\bheading{Comparison \& Qualitative Analysis.}
Prior work can fail for long-context inputs.
For the above example in \Cref{fig:example_health}, we test previous state-of-the-art output2text~\cite{zhang2024extracting} pretrained on \llama-2 and unbridle the maximum sequence to 4,096.
The output is ``How do you manage the tension and tension that is causing you to go crazy?'', which is completely different.
On the contrary, ours only misses certain tokens.

We found that the missed tokens are synonym of the ground truth tokens and have similar input embeddings.
Take the first missed token in \Cref{fig:example_health} as an example, the ground truth token is ``friend'' while the inversion is ``boyfriend'' because its embedding is the most similar to the inverted input embedding.
After a manual checking, we found the embedding of the ground truth token ``friend'' is the third most similar token (which is apparently similar to the embedding of ``boyfriend''), thus is missed during token generation.
That is why smaller learning rates could benefit the inversion because of more refined updating.
One potential improvement could be beam searching to cover all the possible paths.

\iheading{Learning Rates \& Distance Metric.}
To begin with, we investigate different learning rates and distance metrics of our TBS attack in \Cref{fig:wb-long-lr-loss-dataset}.
The results are expected because the self-attention leverages inner product to compute attention and subsequent \acp{is}, which makes COS more reliable distance metric than MSE.
However, computing COS requires higher VRAM because the matrix computation has space complexity $O(n^2)$ for a $n$-length prompts.
Besides, as mentioned earlier, we found smaller learning rates improve the inversion on deep layers.
The potential reason can be more fine-grained input embedding inversion, which leads to more accurate candidate token recovery.
Next, we explore to improve the attack performance by additional settings.

\iheading{Improved Attack Settings.}
\Cref{tab:dm_penalty_and_basis} shows inversion with application of our token distribution matching penalty and usage of SVD singular basis.
We make two observations: 1) the penalty can improve the inversion at larger learning rate 0.0005 which may be useful to accelerate inversion through faster convergence (\ie, fewer steps).
2) usage of SVD singular basis can also improve the inversion performance regardless of learning rates, possibly because it enables more precise input embedding inversion than unbiased basis.
Nevertheless, SVD singular basis may not work if applying optimization-based attack to derivatives under the black-box setting (see \Cref{subsec:black-box-results}).

\iheading{Larger Models.}
Our white-box attacks are size-agnostic thus can scale to larger models, at the expense of higher VRAM cost and optimization time.
As shown in the \Cref{tab:dm_penalty_and_basis}, our TBS attack remains effective for \llama-3.3-70B on Mentalhealth.
Notably, compared to smaller 7B \llama model, the inverted texts are closer to the ground truth as indicated by higher CS and F1 scores, possibly because of more information retained in wider \acp{is} (\ie, 8192 for 70B and 4096 for 7B).
However, as the TBS attack requires forwarding and backpropagation to iteratively update the inverted input, larger models can significantly increase the optimization time and GPU memory.
For example, attacking 70B is 7 times slower and costs 10 times more memory than 7B.

\iheading{Input Data Type.}
We observe that the input data type can influence the inversion performance.
In \Cref{fig:wb-long-lr-loss-dataset}, the coding data are more difficult to be inverted comparing to medical texts.
Notably, the F1 score on coding datasets (92.05 and 94.50) are slightly lower than healthcare dialogue data under the optimal attack setting.
After manual checking of inverted inputs, we found that the additional errors appear in the description of coding prompts instead of the main code.
This may indicate that the model knowledge may also influence our optimization-based inversion attack because of the attention assignment to tokens of different topics is unequal.
Therefore, in the following we investigate whether the coding-enhanced model can lead to higher inversion risk.

\begin{table}[t]
\caption{Evaluation of different TBS attack settings for \llama-3 model.}
\label{tab:dm_penalty_and_basis}
\resizebox{\linewidth}{!}{
\begin{tabular}{@{}ccccccccc@{}}
\toprule
\textbf{Dataset} & \textbf{Size} & \textbf{Basis} & \textbf{Penalty} & \textbf{$\mu$} & \textbf{CS} & \textbf{BLEU} & \textbf{ROUGE} & \textbf{F1} \\ \midrule
\multirow{6}{*}{EC} & \multirow{6}{*}{8B} & \multirow{2}{*}{Unbiased} & \multirow{2}{*}{0} & 1e-4 & 95.49$\pm$ 0.59 & 64.93$\pm$ 2.90 & 0.90 & 76.22 $\pm$ 2.11 \\ \cmidrule(l){5-9} 
 &  & &   & 5e-4 & 88.79$\pm$ 0.83 & 27.06$\pm$ 2.04 & 0.66 & 49.56 $\pm$ 1.65 \\ \cmidrule(l){3-9} & 
 & \multirow{2}{*}{Unbiased} & \multirow{2}{*}{1e-3} & 1e-4 & 94.73$\pm$ 1.28 & 58.09$\pm$ 6.56 & 0.87 & 72.48 $\pm$ 3.70 \\ \cmidrule(l){5-9} & 
 &  &  & 5e-4 & 90.82$\pm$ 1.05 & 34.20$\pm$ 3.03 & 0.71 & 54.70 $\pm$ 2.29 \\ \cmidrule(l){3-9} & 
 & \multirow{2}{*}{SVD} & \multirow{2}{*}{0} & 1e-4 & 95.87$\pm$ 0.95 & 64.34$\pm$ 3.05 & 0.88 & 73.90 $\pm$ 2.60 \\ \cmidrule(l){5-9} 
 &  &  & & 5e-4 & 98.20$\pm$ 0.49 & 72.09$\pm$ 3.14 & 0.91 & 80.73 $\pm$ 2.22 \\ \midrule
\multirow{6}{*}{MH} & \multirow{6}{*}{8B} & \multirow{2}{*}{Unbiased} & \multirow{2}{*}{0} & 1e-4 & 95.02$\pm$ 0.70 & 80.36$\pm$ 1.47 & 0.95 & 90.63 $\pm$ 0.75 \\ \cmidrule(l){5-9} & 
 &  &  & 5e-4 & 88.59$\pm$ 1.66 & 56.69$\pm$ 3.04 & 0.87 & 78.53 $\pm$ 1.99 \\ \cmidrule(l){3-9} & 
 & \multirow{2}{*}{Unbiased} & \multirow{2}{*}{1e-3} & 1e-4 & 94.79$\pm$ 0.77 & 77.55$\pm$ 2.00 & 0.94 & 88.92 $\pm$ 1.20 \\ \cmidrule(l){5-9} & 
 &  &  & 5e-4 & 89.09$\pm$ 0.79 & 59.03$\pm$ 2.25 & 0.88 & 80.30 $\pm$ 1.08 \\ \cmidrule(l){3-9} & 
 & \multirow{2}{*}{SVD} & \multirow{2}{*}{0} & 1e-4 & 97.64$\pm$ 0.39 & 86.70$\pm$ 1.85 & 0.96 & 93.27 $\pm$ 1.05 \\ \cmidrule(l){5-9} 
 &  &  &  & 5e-4 & 98.00$\pm$ 0.57 & 90.24$\pm$ 1.24 & 0.97 & 94.99 $\pm$ 0.61 \\ \cmidrule(l){2-9} 
 & \multirow{1}{*}{70B} & \multirow{1}{*}{Unbiased} & \multirow{1}{*}{0} & 5e-4 & 95.80$\pm$ 1.17 & 74.17$\pm$ 9.44 & 0.94 & 88.34 $\pm$ 5.17 \\
 
 \bottomrule
\end{tabular}
}
\end{table}

\bheading{Evaluation of Domain-specialized Models.}
We consider \qwen2.5-Coder as the coding-specialized model because of its high ranking on the leaderboard~\cite{codemodelleaderboard}.
Here we evaluate the top 50 longest coding prompts from two benchmarks.
We also evaluate \qwen2.5 as a general-purpose model.
\Cref{tab:tbs-qwen-results} presents the results on two coding benchmarks.
First, we observe that inversion is significantly better: 
the EM scores are around 50\% for the top long prompts of both benchmarks and the inverted tokens F1 scores are around 99\%.
Second, compared to the general-purpose model \qwen2.5, the coding-specialized model \qwen2.5-Coder has \acp{is} more susceptible to inversion attack because of higher EM rates.  

Surprisingly, we observe that the \acp{is} of \qwen2.5 is easier to invert than \llama-3 models.
We hypothesize no impact from their pretraining data and investigate the architecture difference.
We exclude the causes from model width and trainable parameters up to the middle layer because they are similar for two models: \qwen2.5 is of width 3,584 and 3.8B parameters while \llama-3 is of width 4,096 and 4B parameters.
After carefully examining their implementations, we found the main difference lies in the attention module: \qwen2.5 applies QKV bias~\cite{qkvbias} while \llama-3 does not by default.
Therefore, we suspect that attention bias may amplify the feature representation of \acp{is} thus, as a side effect, enable better inversion.
Due to the huge cost of pretraining to obtain similar bias for \llama-3, we leave more detailed impact analysis of attention bias for future work.

\lesson{On long-context inputs, our optimization-based attack TBS can invert nearly all tokens in the correct order to preserver semantics.}

\begin{table}[t]
\caption{Evaluation results of \qwen2.5 and \qwen2.5-Coder.}
\label{tab:tbs-qwen-results}
\resizebox{\linewidth}{!}{
\begin{tabular}{@{}cccccccc@{}}
\toprule
\textbf{Dataset} & \textbf{Model} & \textbf{Distance} & \textbf{CS} & \textbf{BLEU} & \textbf{ROUGE} & \textbf{EM} & \textbf{F1} \\ \midrule
\multirow{3}{*}{EC} & Qwen2.5 & MSE &99.03$\pm$ 0.26 & 98.90$\pm$ 0.18 & 1.00 & 12.90$\pm$ 0.06 & 98.90 $\pm$ 0.21  \\ \cmidrule(l){2-8} 
 & \begin{tabular}[c]{@{}c@{}}Qwen2.5-\\ Coder\end{tabular} & MSE & 99.78$\pm$ 0.12 & 99.60$\pm$ 0.13 & 1.00 & 64.00$\pm$ 0.07 & 99.80 $\pm$ 0.06 \\ \cmidrule(l){2-8} 
 & \begin{tabular}[c]{@{}c@{}}Qwen2.5-\\ Coder\end{tabular} & COS & 99.96$\pm$ 0.03 & 99.12$\pm$ 0.34 & 1.00 & 45.00$\pm$ 0.11 & 99.47 $\pm$ 0.14 \\ \midrule
\multirow{2}{*}{CA} 
 & \begin{tabular}[c]{@{}c@{}}Qwen2.5-\\ Coder\end{tabular} & MSE & 99.86$\pm$ 0.08 & 99.78$\pm$ 0.05 & 1.00 & 50.00$\pm$ 0.07 & 99.82 $\pm$ 0.04 \\ \cmidrule(l){2-8} 
 & \begin{tabular}[c]{@{}c@{}}Qwen2.5-\\ Coder\end{tabular} & COS & 99.72$\pm$ 0.28 & 99.46$\pm$ 0.37 & 1.00 & 40.00$\pm$ 0.16 & 99.76 $\pm$ 0.08 \\ \bottomrule
\end{tabular}
}
\end{table}

\subsection{Black-box Inversion Attack}
\label{subsec:black-box-results}

In this section, we evaluate and compare our replication-based and generation-based black-box inversion attacks.

\bheading{Attack Settings.}
We use the prompts from Mentalhealth and Evolcode for test. 
For both black-box attacks, we assume the adversary uses NoRobots~\cite{no_robots} as training data for replication model and inversion model training.
NoRobots is composed of 9.5k+ supervised finetuning samples from general topics, which aligns with our assumption that the adversary has data of similar distribution.
We consider \llama-3 owned by the adversary and target \llama-3-Doctor and \llama-3.1 as two black-box deployed models.
In this section we also test \acp{is} of the middle layer as above.

\begin{figure}[t]
    \centering
    \includegraphics[width=0.9\linewidth]{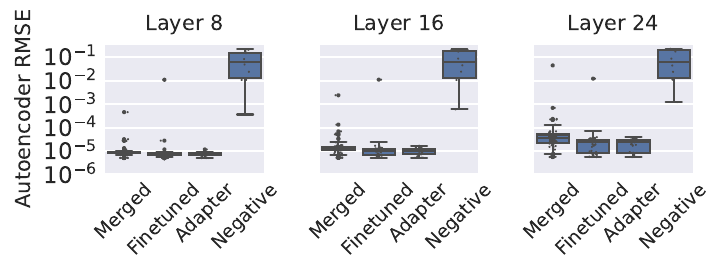}
    \caption{The distribution of autoencoder's reconstruction error to detect derivatives of \llama-3.}
    \label{fig:boxplot-autoencoder-rmse}
\end{figure}

\begin{figure}[t]
    \centering
    \includegraphics[width=0.99\linewidth]{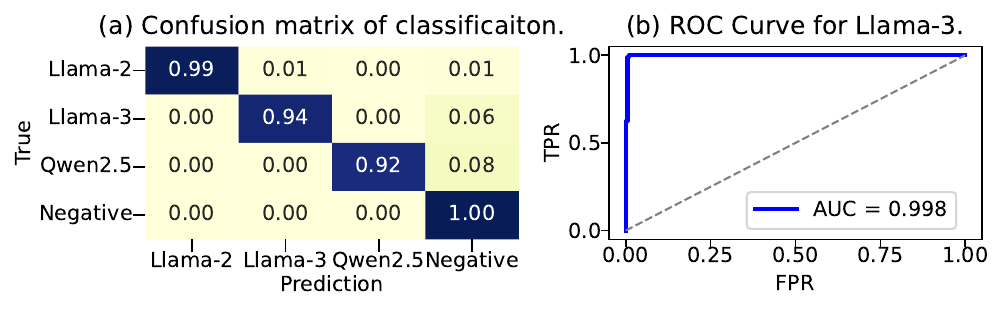}
    \caption{Evaluation of model type identification.}
    \label{fig:ae-detect}
\end{figure}

\bheading{Model Type Identification.}
Given unknown \acp{is}, the adversary first identifies the deployed model type.
To test the identify model type, we trained and crawled from Hugging Face 75 finetuned models, 70 merged models and 56 adapters of most downloading of \llama-3.
In addition, we select 13 \acp{llm} independent models (\ie, those not derived from \llama-3) as negatives, which are listed in the \Cref{app:additional_results}.

\Cref{fig:boxplot-autoencoder-rmse} demonstrates the distribution of autoencoder's Reconstruction MSE (RMSE) on three tested layers.
For each tested layer, we use the test dataset of 500 samples of NoRobots as the probing data, and train the 3-layer autoencoder for 10 epochs.
We found that most derivatives have separate error ranges to the independent models, which enables binary classifications by thresholding.
Note that there are a small number of outliers that can be identified as independent model (False Negative).
We check the false negatives and found they are caused by labeling errors based on crawled model name.
Therefore, in the following test, we check the top downloading derivatives to remove potential errors.

To evaluate the identification of ensemble autoencoders, we additionally crawl the derivatives of \llama-2 and \qwen2.5 in a similar manner and ensemble the autoencoders of middle-layer \acp{is} of three \acp{llm}.
\Cref{fig:ae-detect} exhibits the confusion matrix of model type classification (left) and the ROC curve with AUC score (right).
We found our ensemble autoencoder can almost perfectly classify the target mode type.
Therefore, the adversary can leverage the publicly available base \ac{llm} to apply the model replication-based inversion.

\begin{table}[t]
\caption{Results of the replication-inversion attack.}
\label{tab:replic-results}
\resizebox{\linewidth}{!}{
\begin{tabular}{@{}ccccccccc@{}}
\toprule
\multicolumn{1}{c}{\textbf{Test Dataset}} & \textbf{Attack} & \textbf{Target} & Basis & \textbf{lr} & \textbf{CS} & \textbf{BLEU} & \textbf{ROUGE} & \textbf{F1} \\ \midrule
\multirow{10}{*}{MH} & \multirow{6}{*}{Transferred} & Doctor & \multirow{2}{*}{SVD} & \multirow{2}{*}{1e-4} & 52.87$\pm$ 1.30 & 0.98$\pm$ 0.25 & 0.22 & 14.79 $\pm$ 1.77 \\ \cmidrule(lr){3-3} \cmidrule(l){6-9} 
 &  & Llama-3.1 &  &  & 56.21$\pm$ 1.24 & 1.85$\pm$ 0.47 & 0.29 & 19.33 $\pm$ 1.66 \\ \cmidrule(l){3-9} 
 &  & \multirow{2}{*}{Doctor} & \multirow{4}{*}{Unbiased} & 1e-4 & 64.50$\pm$ 2.70 & 6.64$\pm$ 1.62 & 0.45 & 32.12 $\pm$ 3.05 \\ \cmidrule(l){5-9} 
 &  &  &  & 5e-4 & 58.36$\pm$ 1.12 & 0.90$\pm$ 0.08 & 0.23 & 17.91 $\pm$ 0.54 \\ \cmidrule(lr){3-3} \cmidrule(l){5-9} 
 &  & \multirow{2}{*}{Llama-3.1} &  & 1e-4 & 61.19$\pm$ 2.14 & 4.72$\pm$ 0.76 & 0.40 & 27.53 $\pm$ 1.85 \\ \cmidrule(l){5-9} 
 &  &  &  & 5e-4 & 63.60$\pm$ 2.10 & 2.45$\pm$ 0.51 & 0.30 & 22.10 $\pm$ 1.35 \\ \cmidrule(l){2-9} 
 & \multirow{4}{*}{\begin{tabular}[c]{@{}c@{}}Replicated\\ (Ours)\end{tabular}} & \multirow{2}{*}{Doctor} & \multirow{4}{*}{Unbiased} & 1e-4 & 72.32$\pm$ 2.64 & 16.33$\pm$ 2.23 & 0.58 & 43.35 $\pm$ 2.31 \\ \cmidrule(l){5-9} 
 &  &  &  & 5e-4 & 66.74$\pm$ 1.79 & 5.54$\pm$ 1.30 & 0.42 & 30.59 $\pm$ 1.93 \\ \cmidrule(lr){3-3} \cmidrule(l){5-9} 
 &  & \multirow{2}{*}{Llama-3.1} &  & 1e-4 & 68.33$\pm$ 2.33 & 4.91$\pm$ 0.71 & 0.45 & 30.76 $\pm$ 1.01 \\ \cmidrule(l){5-9} 
 &  &  &  & 5e-4 & 61.90$\pm$ 1.57 & 0.87$\pm$ 0.16 & 0.25 & 17.53 $\pm$ 0.82 \\ \midrule
\multicolumn{1}{c}{\multirow{4}{*}{EC}} & \multirow{2}{*}{Transferred} & Doctor & \multirow{2}{*}{Unbiased} & 1e-4 & 62.01$\pm$1.70 & 3.60$\pm$0.68 & 0.27 & 18.56$\pm$1.60 \\ \cmidrule(lr){3-3} \cmidrule(l){5-9} 
\multicolumn{1}{c}{} &  & Llama-3.1 &  & 1e-4 & 63.05$\pm$3.08 & 2.13$\pm$0.46 & 0.23 & 15.63$\pm$1.60 \\ \cmidrule(l){2-9} 
\multicolumn{1}{c}{} & \multirow{2}{*}{\begin{tabular}[c]{@{}c@{}}Replicated\\ (Ours)\end{tabular}} & Doctor & \multirow{2}{*}{Unbiased} & 1e-4 & 76.92$\pm$2.31 & 9.71$\pm$1.63 & 0.47 & 29.60$\pm$2.17 \\ \cmidrule(lr){3-3} \cmidrule(l){5-9} 
\multicolumn{1}{c}{} &  & Llama-3.1 &  & 1e-4 & 62.22$\pm$2.00 & 1.91$\pm$0.43 & 0.26 & 16.74$\pm$1.21 \\ \bottomrule
\end{tabular}
}
\end{table}

\bheading{Replication-based Inversion.}
We use \llama-3 replicate to the target models (ChatDoctor and \llama-3.1) on NoRobots for three epochs with Qlora~\cite{qlora} with learning rate 0.0001.
As for the inversion attack, we use MSE as the loss distance. 
\Cref{tab:replic-results} presents the evaluation of our replication-based inversion on Mentalhealth and Evolcode, from which we can make four findings:
1) The use of SVD singular basis can degrade the attack because of the discrepancy between the inversion and the target input embedding spaces.
2) The replicated model with a low attack learning rate leads to better input inversion because of better input embedding alignment.
3) The post-training can offer better protection against inversion than finetuning, as \llama-3.1 is more difficult 
to invert input information (\eg, fewer matched tokens as indicated by lower F1).
4) The distribution gap between the adversary data and the victim's queries can degrade the inversion.
On coding data, the improvement from model replication is lower than on medical texts, as shown under the optimal setting (ChatDoctor with unbiased basis and 1e-4).

In sum, the optimization-based attack performance is deteriorated by limited attack information in the black-box setting and can be difficult to improve.
As discussed earlier, the token recovery can be sensitive to the cosine difference between the inverted and the ground truth embeddings.
The Transformer and input embedding weights of the derived model enlarge this inverted difference, resulting in worse token recovery.
Next, we evaluate the generative model-based inversion attack which is based on observed \acp{is} as context and can be more robust to the embedding gap.

\begin{table}[t]
\caption{Black-box inversion on short-context input text.}
\label{tab:black-box-short-context}
\resizebox{\linewidth}{!}{
\begin{tabular}{@{}ccccccc@{}}
\toprule
\textbf{Dataset} & \textbf{Average Length} & \textbf{CS} & \textbf{BLEU} & \textbf{ROUGE} & \textbf{EM} & \textbf{F1} \\ \midrule
Instruct-2M & 26.16 & 97.85$\pm$ 0.24 & 92.73$\pm$ 0.58 & 0.97 & 61.38$\pm$ 0.02 & 96.87 $\pm$ 0.27 \\ \midrule
Norobot-test & 89.04 & 83.90$\pm$ 0.84 & 51.34$\pm$ 1.55 & 0.64 & 13.00$\pm$ 0.02 & 65.76 $\pm$ 1.44 \\ \midrule
SyntheticGPT & 177.48 & 93.62$\pm$ 0.61 & 46.98$\pm$ 1.25 & 0.67 & 0.0 & 72.90 $\pm$ 0.88 \\ \bottomrule
\end{tabular}
}
\end{table}

\bheading{Generation-based Inversion.}
As for inversion model, the adversary adopts the recent \texttt{T5-base} (T5)~\cite{raffel2020exploring} as it is one of the most commonly used encoder-decoder model.
We train T5 with learning rate 0.0002 and trains on Instruction-2M and NoRobots for short-context and long-context inputs, respectively.
We also use SyntheticGPT~\cite{zhang2024extracting} for short-context inversion evaluation.
To align with replication-based black-box attack, the deployed model is \llama-3.1 as our target.

\iheading{Inversion of Short-context Input.}
We train inversion model on Instruction-2M for 1 epoch.
\Cref{tab:black-box-short-context} presents the inversion evaluation on short-context datasets.
The attack performs best on the same-distribution test dataset of Instruction-2M and achieves even better performance than our white-box optimization-based inversion attack.
However, the downside for generative model is that it can hardly scale to longer prompts and the performance is susceptible to distribution shift.
As the length increases, the semantic similarity scores drop significantly on longer SyntheticGPT dataset.

\begin{figure}[t]
    \centering
    \includegraphics[width=0.99\linewidth]{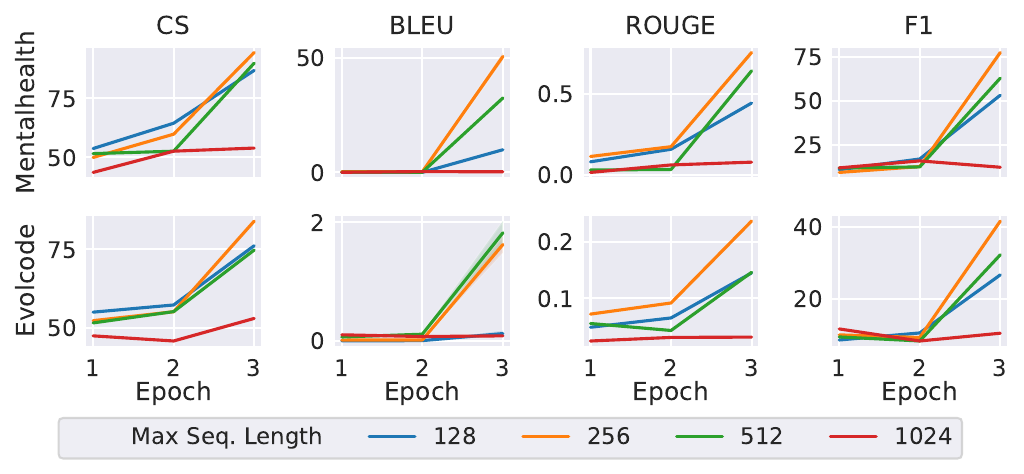}
    \caption{Generative inversion on long-context datasets.}
    \label{fig:black-box-gen}
\end{figure}

\iheading{Inversion of Long-context Inputs.}
To understand the limit of input length for generative inversion, we train with NoRobots that contains long-context prompts of up to 3,384 tokens (detailed length distribution is shown in \Cref{fig:norobot-train-histplot}).
To investigate the impact of training sequence length, we set maximum sequence length to 128, 256, 512 and 1,024.
As NoRobots only contains 9,485 samples, we train inversion model up to 3 epochs that counts for similar model update steps to Instruction-2M for ensuring model convergence.

After training, we evaluate on the top 100 longest prompts from Mentalhealth and Evolcode with greedy sampling of maximum length 4,096 to ensure full generation.
\Cref{fig:black-box-gen} shows the results from which we can make three observations.
First, we found that the maximum sequence length 256 is the optimal setting for NoRobots dataset, where shorter or longer limits can result in worse inversion.
Second, too long sequence limit 1024 can hinder the inversion quality.
Third, the data type plays a central role to accurately invert \acp{is}: the model cannot generate meaningful inverted texts on coding prompts (Evolcode) but achieve similar performance as optimization-based attack on the similar data (Mentalhealth).

\begin{figure}
    \centering
    \includegraphics[width=0.99\linewidth]{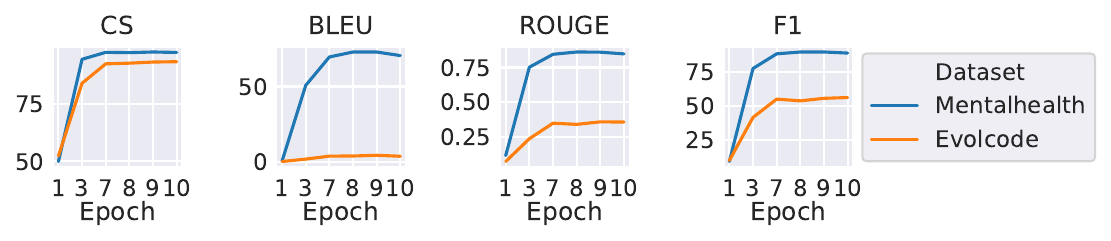}
    \caption{Evaluation of converged inversion models.}
    \label{fig:black-gen-long-epoch}
\end{figure}

\iheading{Longer Training Epochs.}
We also observe that the inversion model only generates prompts until the third epoch, which may indicate that that more epochs can improve the performance. 
In \Cref{fig:black-gen-long-epoch}, we extend the epochs to 10 for model convergence while keeping maximum sequence length 256.
We found that the converged performance on Mentalhealth is lower than the best performance of optimization-based inversion as shown in \Cref{fig:wb-long-lr-loss-dataset} (\eg, 94.99 BLEU and 96.7 F1 on Mentalhealth).
On the other hand, the inversion on coding data does not get improved as medical texts because of the distribution gap, which further highlights importance of the same distribution assumption.
\Cref{subsecapp:failurecase} includes more analysis on failure cases due to the distribution mismatch.

\begin{table}[t]
\caption{Evaluation of generative inversion on 70B model.}
\label{tab:70b-mentalhealth}
\resizebox{\linewidth}{!}{
\begin{tabular}{@{}ccccccc@{}}
\toprule
\textbf{Dataset} & \textbf{Max Seq. Length} & \textbf{CS}     & \textbf{BLEU}   & \textbf{ROUGE} & \textbf{EM}    & \textbf{F1}      \\ \midrule
\multirow{3}{*}{MH} & 256                       & 87.92$\pm$ 0.45 & 21.50$\pm$ 0.57 & 0.50           & 0.0            & 55.56 $\pm$ 0.57 \\\cmidrule(l){2-7}
& 512                       & 97.76$\pm$ 0.22 & 79.39$\pm$ 1.84 & 0.88           & 2.00$\pm$ 0.01 & 90.75 $\pm$ 0.90 \\ \cmidrule(l){2-7}
& 1024                      & 97.44$\pm$ 0.23 & 76.26$\pm$ 1.82 & 0.87           & 1.00$\pm$ 0.01 & 88.72 $\pm$ 0.90 \\ 
\midrule
\multirow{3}{*}{EC} & 256 & 74.85$\pm$ 0.81 &  3.69$\pm$ 0.22&  0.16 	&  0.00 &  28.52 $\pm$  0.89 \\ \cmidrule(l){2-7}

& 512 & 94.18$\pm$ 0.42 & 	 20.98$\pm$ 0.67 & 	 0.42 	 & 0.0 & 	 60.44 $\pm$ 1.15 \\ \cmidrule(l){2-7}

& 1024  &  91.74$\pm$ 0.55 & 	 18.08$\pm$ 0.81 & 	 0.38&  	 0.0 & 	 55.62 $\pm$ 1.26 \\
\bottomrule
\end{tabular}}
\end{table}
\iheading{Larger Models.}
\Cref{tab:70b-mentalhealth} validates the effectiveness of the generative inversion on \llama-3.3-70B model.
Compared to previous smaller \acp{llm}, the optimal maximum sequence length for inversion of 70B models is doubled: from 256 to 512.
In particular, there are even exact matching cases (\eg, inversion trained with maximum sequence length 512).
We observe that, despite deeper and wider \acp{is}, more input tokens can be inverted as signified by higher F1 scores and CS scores.
This result aligns with our previous white-box inversion evaluation on the 70B model.

\iheading{Comparison with Optimization-based Inversion.}
We examine the generated inversion inputs and qualitatively compare them with those optimized by our white-box attack.
Although both achieve similar semantic similarity and token matching score, the reason for failed inversion tokens is different.
As mentioned previously, the optimized inversion inputs may introduce or replace the true token with noisy tokens of different languages or Unicode which hinders the readability but also partially reveals the meaning of true inputs.
On the other hand, generative inverted texts contain no noisy tokens but may miss entirely some sentences, especially those in the middle or the end of the original input.
This is a disadvantage to the adversary because of potential key information loss.

\iheading{Transferability.}
We explore whether the generative inversion model can directly transfer to the derived target model.
We evaluate the T5 model trained on \llama-3.1's \acp{is} and evaluate on \llama-3 (see \Cref{tab:gen-blackbox-transfer-llama3}).
The inversion performance is close to the non-transfer case as shown in \Cref{fig:black-box-gen}.
This validates our claim that the generation-based can better tolerate the embedding gap than the optimization-based attack.

\begin{table}[t]
\caption{Transferability of generative inversion to Llama-3.}
\label{tab:gen-blackbox-transfer-llama3}
\resizebox{\linewidth}{!}{
\begin{tabular}{@{}cccccc@{}}
\toprule
\textbf{Dataset} & \textbf{Max. Seq. Length} & \textbf{CS} & \textbf{BLEU} & \textbf{ROUGE} & \textbf{F1} \\ \midrule
\multirow{3}{*}{Mentalhealth} & 128 & 86.67$\pm$ 0.43 & 10.35$\pm$ 0.66 & 0.45 & 54.06 $\pm$ 0.68 \\ \cmidrule(l){2-6} 
 & 256 & 94.53$\pm$ 0.30 & 53.12$\pm$ 1.71 & 0.77 & 78.56 $\pm$ 0.74 \\ \cmidrule(l){2-6} 
 & 512 & 90.03$\pm$ 0.36 & 32.71$\pm$ 1.05 & 0.64 & 63.32 $\pm$ 0.62  \\ \midrule
\multirow{3}{*}{Evolcode} & 128 & 76.97$\pm$ 0.81 & 0.13$\pm$ 0.05 & 0.15 & 28.37 $\pm$ 1.01  \\ \cmidrule(l){2-6} 
 & 256 & 84.04$\pm$ 0.76 & 2.10$\pm$ 0.19 & 0.23 & 40.63 $\pm$ 1.19  \\ \cmidrule(l){2-6} 
 & 512 & 74.78$\pm$ 0.83 & 1.57$\pm$ 0.16 & 0.15 & 31.45 $\pm$ 1.00  \\ \bottomrule
\end{tabular}
}
\end{table}

\subsection{Defense Evaluation}
\label{subsec:defense}

In this section, we test four practical defenses including \ac{dp}, pruning, quantization, dropout and noisy input embedding.
In \Cref{sec:discussion}, we discuss and provide mitigation suggestions.

\begin{figure*}
    \centering
    \includegraphics[width=0.95\linewidth]{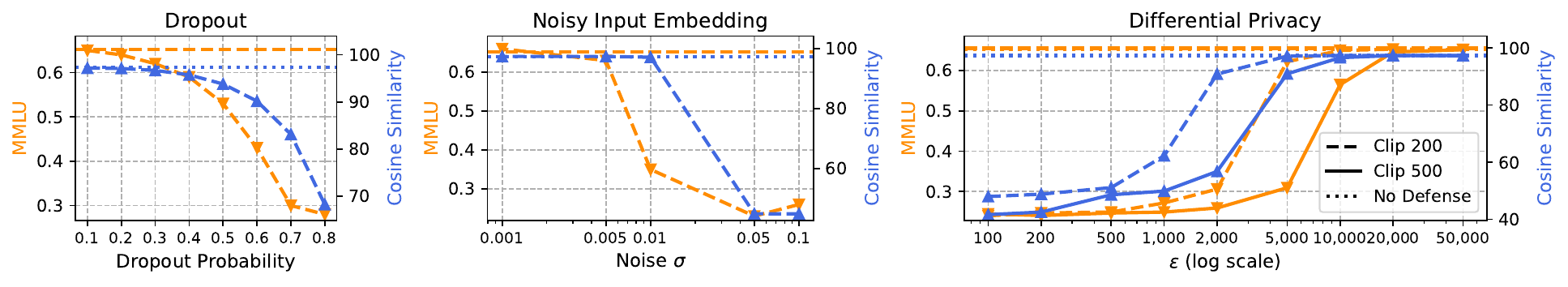}
    \caption{Evaluation of defenses including noisy input embedding, dropout and differential privacy with Laplace mechanism.}

    \label{fig:defendtradeoff}
\end{figure*}
\bheading{Setting.}
We use Mentalhealth and consider the collaborative inference server who receives the \acp{is} $\interstates{\llm_\cV}{l}$ at the middle layer $l=16$ for \llama-3-8B.
The client applies defenses before the release of \acp{is} for the first-round inference.
For attack, we test the generation-based inversion attack trained for 10 epochs on NoRobots with the optimal maximum sequence length 256.
In terms of utility, we evaluate the model with five-shot Massive Multitask Language Understanding (MMLU)~\cite{hendrycks2021measuring}, which refers to accuracy on the benchmark questions.
We fix the random seed to ensure reproducibility for defenses involving randomness.

\bheading{Quantization.}
We first test the 4-bit and 8-bit model quantization which, as a common technique to reduce memory cost, can alter the adversary-observed \acp{is} through through forward-propagated error to degrade inversion performance.
We test the model quantization and present the results in \Cref{tab:quantization-defense} of \Cref{app:additional_results}.
We note that only 4-bit slightly effect the MMLU but the inversion quality remains nearly unaffected, indicating that quantization cannot defend our attack.

\bheading{Dropout.}
We apply the dropout on \acp{is} of probability $p\in\{0.1,\cdots, 0.8\}$ that nullifies $p$ elements and scales the rest by $1/(1-p)$.
As $p$ increases, the MMLU score will firstly decrease.
Meanwhile, we can see that MMLU drops faster than the CS score.
Among the highest dropout probabilities we tested $p\geq 0.7$, the MMLU score is close to the random guess ($\sim0.25$) while we can still achieve inversion of CS score higher than 70.
This indicates that dropout cannot fail our generative inversion while preserving the utility.

\bheading{Noisy Input Embedding.}
The defender can blur the input embedding to obfuscate the \acp{is}.
Therefore, we add the noise following Gaussian distribution of variance $\sigma^2$, and show the evaluation results in the middle of \Cref{fig:defendtradeoff}.
For low noise scale $\sigma \leq 0.005$, there is negligible affect over the inference and inversion.
However, we observe that $\sigma=0.01$ is a turning point: the MMLU score become lower than 0.4 while the inversion is almost unaffected. 
Our attack can be effectively defended at high noise scale $\sigma\in\{0.05,0.1\}$, but the MMLU also decreases to the random guess level.
To wrap up, noisy input embedding cannot achieve a perfect trade-off between the inference utility and \acp{is} privacy.

\bheading{Differential Privacy.}
We add Laplacian noise to the \ac{is} to achieve $\epsilon$-\ac{dp}: $\delta\sim\text{Lap}(0, \epsilon / \Delta_s)$, where $\Delta_s$ is the sensitivity.
As the maximum value of \acp{is} is not bounded, we clip the $\interstates{\llm_\cV}{l}$ by $C_{\Delta_s}\in\{200, 500\}$, because we observe the highest \acp{is} are within this range on Mentalhealth.
This makes $\Delta_s = 2C_{\Delta_s}$.

The rightmost figure of \Cref{fig:defendtradeoff} shows the inversion performance (measured by CS) and model utility on different $\epsilon$ and two clipping cases, where the horizontal dotted lines represent no \ac{dp} protection.
We observe that the blue curves (inverted text similarity) increase earlier than orange curves (model utility) in both clipping cases.
This indicates that the adversary can gain advantage with low $\epsilon$ that almost nullifies the model.
In other words, with mild drop of model utility (\eg, $\epsilon=5,000$ for $C_{\Delta_s}=200$), our generation-based attack remains equally effective.

\lesson{Our generation-based inversion can achieve accurate inversion on data of similar distribution and bypass practical defenses directly applied on \acp{is}.}

\section{Related Work}
\label{sec:relatedwork}

In this section, we review applications of \ac{is} and inversion attacks in \ac{nlp}.

\bheading{Application of Internal States.}
Recently, several works have shown that the \ac{is} are strong indicator of hallucination factual error and safety status, which provide a practical strategy for \ac{llm} holder and regulator to surveil the model behavior.
\citeauthor{azaria-mitchell-2023-internal} first discovered that the \acp{is} can indicate factual errors and found that deep layer's \acp{is} can train accurate hallucination detector~\cite{azaria-mitchell-2023-internal}.
The \acp{is}-based hallucination detection can also be achieved by eigenvalues~\cite{chen2024inside}, unsupervised clustering~\cite{su-etal-2024-unsupervised} and linear probing~\cite{DBLP:journals/corr/abs-2406-15927} to stop the mistakes on the fly~\cite{ye2024physics}.
The middle layer has \acp{is} that can indicate malicious and benign queries and can be used as safety layer to defend jailbreak attacks~\cite{li2024safety}.
Additionally, \acp{is} can reveal latent knowledge~\cite{burns2023discovering}, membership privacy of query data~\cite{ji2024llm}, emotion~\cite{zhou-etal-2024-alignment,zou2023representation}, internal symbolic calculations~\cite{chen2024states} and potential backdoors~\cite{lamparth2024analyzing,dong2024trojaningplugins}.
As more applications of \acp{is} emerge, the potential privacy risk should be carefully examined and our work takes the first step to testify the \acp{is} inversion feasibility.

\bheading{Inversion Attacks.}
In addition to the embedding~\cite{morris-etal-2023-text} and outputs~\cite{morris2024language,zhang2024extracting}, the risk of data inversion also exists for shared gradient in decentralized learning~\cite{DBLP:conf/nips/LiYLWHYX23,chen2024unveiling_ccs24}, unlearning~\cite{sp24unlearninversion}, outsourced shallow-layer inference~\cite{zheng2023input} and KV cache~\cite{yang2024lookefficientsecureondevice} which are the most related works to ours.
Nevertheless, our methods can extend to middle or last layers and can generalize to KV cache.
Note that the side-channel attack to \ac{llm} serving~\cite{wu2025promptleakkvcache} is also similar to our inversion, but leverages timing difference in accelerated inference.
There is also finetuning-based defenses~\cite{dpforward_ccs23} but is limited to small \acp{lm} because of performance drop for \acp{llm}.
Recently, there are concurrent works~\cite{pia_coll_sp25,luo2025prompt} attempting to invert prompt from the malicious server in collaborative learning.
As for comparison, our work includes more comprehensive attacks (white-box and black-box) and has validated the attack effectiveness on long inputs (4k+ tokens) and large models (up to 70B).

\section{Discussion \& Conclusion}
\label{sec:discussion}
We conclude with discussion and future work.

\bheading{Practical Mitigation.}
As we witnessed in \cref{subsec:defense}, directly protecting \acp{is} is still limited to simultaneously offer both model utility and privacy. 
The fundamental reason is that the inference of subsequent layers depends on meaningful \ac{is} values, hence the noise-induced representation obfuscation must preserve sufficient utility by maintaining semantic coherence.
To mitigate privacy leakage caused by \acp{is}, it is essential to safeguard the entire model rather than concentrating solely on individual layers.
One possible direction can be the exploitation of cryptographic tools or confidential computing.
The homomorphic encryption is a promising solution, but current implementations generate unacceptable computational overhead for deployment~\cite{pia_coll_sp25}.
Another countermeasure that can be immediately taken is exploiting the confidential virtual machine in CPU (\eg, AMD SEV~\cite{sev2020strengthening}) and GPU (\eg, H100~\cite{h100-whitepaper}) which can provide confidentiality guarantee and allow addition obfuscation schemes inside the enclaves to hinder side channel attacks.

\bheading{Architecture-based Mitigation.}
We suspect the equal-width architecture design may cause our privacy inversion attacks.
Typically, the layer width in conventional classification models decreases with depth, enabling the compression of input data and contributing to the emergence of Harmless Space (HS)~\cite{chen2024seeing}.
In contrast, Transformer-based language models have a uniform layer width.
This architectural difference may explain why our attack methods are successful.
Thus, a potential architecture-based defense could involve a model with varying sizes of input embeddings, intermediate states, and output embeddings, which, by exploiting information loss, could increase the difficulty of accurate inversion.

\bheading{Unavailable Ground Truth.}
In \Cref{sec:threatmodel}, the adversary aims to recover the exact input, but original inputs are generally unavailable in practice, which can make it infeasible to check the inversion correctness.
In practice, for the open-source models, the adversary can test with surrogate inputs as reference to ensure constraint.
For closed-sourced models, the adversary can collect input data from their interested distributions for querying and training the inversion model.
The convergence can guarantee the inversion for in-distribution data.

\bheading{Limitation.}
Our optimization-attack is sensitive to the hyperparameter and added noises.
For example, TBS cannot recover meaningful inputs even under the highest $\epsilon$ we tested because of noise sensitivity.
Future work can enhance this attack by denoising on \ac{dp}-protected \acp{is}.
Another limitation is the linear complexity (\ie, $O(E)$) during optimization, which can cause longer attack time for larger models.
Also, it is possible to adopt random initialization for optimization-based attacks to avoid local optima. 
\Cref{subsecapp:failurecase} includes more detailed analysis of failure cases and boundary conditions.
To improve the attack, one of the future direction is to come up with black-box inversion attack that can also be as context-free and length-free as the white-box optimization-based attack.

\section{Ethical considerations}
\label{sec:ethical}

Our attacks exploit \acp{is} of models to invert nearly original sensitive input, posing significant threats to user privacy and data security. As our research do not involve human participants and only use public medical dialogue, Mentalhealth and coding datasets, the IRB of authors' intuition, after our consultation, determined that our research does not require further review.
We acknowledge that exemption from IRB review may not fully address all ethical considerations, and have therefore taken the following steps.

\bheading{Ensuring no harm is caused to stakeholders.}
The potential stakeholders include LLM user and LLM service provider and broader privacy community. To mitigate potential harms to individuals, by following existing privacy work, we only select public datasets that are highly-downloaded and appropriately anonymized. In addition, we also manually checked all datasets used in our research to ensure there is no sensitive information (\eg, PII) exposed in our research. We did not attack real-world collaborative inference systems, so no individual privacy is leaked by our research. As for the models, we only used opensource models including \llama, \qwen, and T5 in our work for compliance with model-use license.

\bheading{Responsible disclose to service providers.}
We also share our findings with main cloud LLM service providers that might deploy collaborative inference including OpenAI, Meta AI and Qwen. We recognize that this cannot perfectly mitigate privacy risks, because our attacks, if known to the public, may still be misused by the actual collaborative inference server or IS auditor to recover the user inputs. However, we believe the benefits of publicizing our attacks outweight the potential harm. As our work demonstrates the potential privacy risks of LLM ISs, it will draw increased attention to, not only the ISs inversion threat, but also the general privacy concerns surrounding LLM service systems. This will encourage the LLM practitioners to proactively implement well-established solutions like confidential virtual machine within their systems to offer better privacy protection.

\section{Open Science}

In compliance with the USENIX Security’s Open Science policy, we commit to publicly releasing the source code to implement the attacks, pretrained inversion models on non-sensitive data in our study upon acceptance of this paper and inversion logs of main results. As the datasets used in our paper are all downloadable from Hugging Face, in our artifact we direct users to the datasets downloading link and provide processing scripts. We also provide instructions on how to test our attacks on user's own data, detailed configuration, program scripts, hyperparameters and hardware requirements.
Our code is released at \url{https://doi.org/10.5281/zenodo.15605325}.

\section*{Acknowledgments}
We thank our shepherd and anonymous reviewers for insightful feedback.
We also thank Yiming Wang for insightful discussions that inspired this project.
The work has been supported in part by the National Natural Science Foundation of China (62325207, 62132013, 62302298).
Shaofeng Li is supported by the Start-up Research Fund of Southeast University (No. RF1028624178).

\bibliographystyle{unsrtnat}
\bibliography{ref}

\appendix

\section{Additional Results}

\label{app:additional_results}

\begin{figure}
    \centering
    \includegraphics[width=0.99\linewidth]{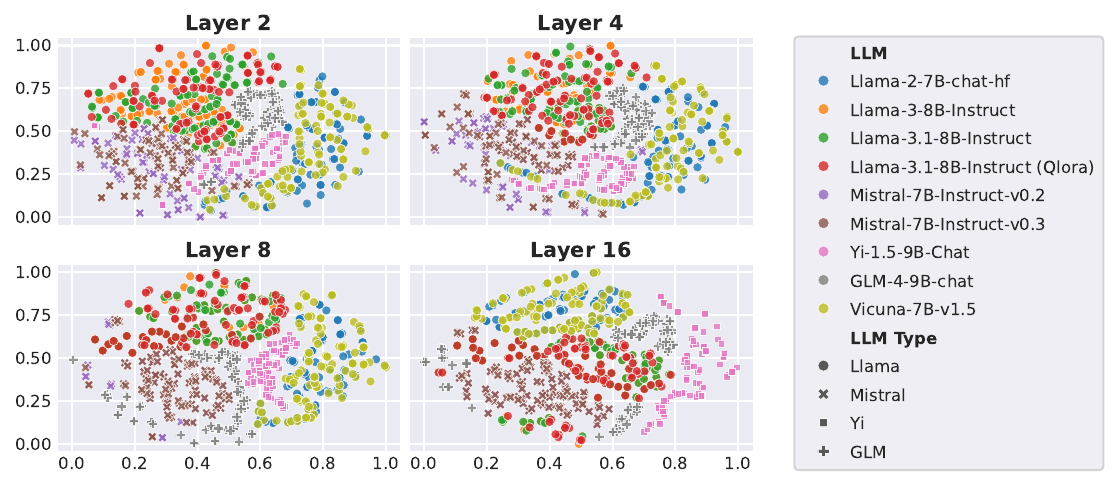}
    \caption{Visualization of internal states for mainstream \acp{llm}. The selected \acp{llm} have the same width 4096 for dimension reduction using t-SNE.}
    \label{fig:tsne-llm}
\end{figure}

\bheading{Visualization of Internal States.}
We applies the t-SNE onto the \acp{is} of various open-source \acp{llm} across different scales.
As visualized in \Cref{fig:tsne-llm}, \acp{llm} rooted from a common pretrained \ac{llm}, have similar \acp{is}.
For example, \llama-3 and \llama-2 are two pretrained models and have separable \acp{is}.
That said, it is possible to train classify the target \ac{llm} type simply using the \acp{is}.
To evaluate model type identification of \llama-3, we select the total 13 \acp{llm} open-sourced before or after \llama-3:Vicuna-7B-v1.5, GLM-4-9B-chat, Qwen2.5-14B-Instruct, Meta-Llama-3.1-8B-Instruct, Qwen2.5-7B-Instruct, Qwen2.5-3B-Instruct, Meta-Llama-3.1-8B (Qlora), gemma-2-9B-it, Llama-3.2-3B-Instruct,  Yi-1.5-9B-Chat, Mistral-7B-Instruct-v0.3, Llama-2-7b-chat-hf, and Mistral-7B-Instruct-v0.2.

\begin{figure}
    \centering
    \includegraphics[width=0.99\linewidth]{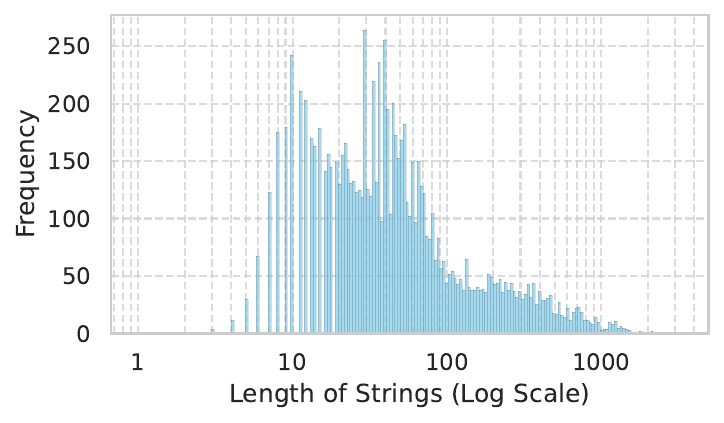}
    \caption{Distribution of token length of \norobots.}
    \label{fig:norobot-train-histplot}
\end{figure}

\bheading{Token length of \norobots.}
\Cref{fig:norobot-train-histplot} shows the histogram of token length of \norobots.
Compared to Instruction2M, we note there are certain long-context training samples which can be indispensable to train long-context inversion model, at expense of increased training time and memory usage.

\bheading{Quantization Defense.}
\Cref{tab:quantization-defense} shows the evaluation of our generative inversion on quantized model.
We use \texttt{bitsandbytes} for \ac{llm} quantization.

\begin{table}[t]
\centering
\caption{Evaluation of quantization as defense.}
\label{tab:quantization-defense}
\resizebox{0.6\linewidth}{!}{
\begin{tabular}{@{}cccc@{}}
\toprule
\textbf{Defense} & \textbf{MMLU} & \textbf{CS}     & \textbf{F1}                        \\ \midrule
No Quant         & 0.65          & 97.33$\pm$ 0.25 & 88.77 $\pm$ 0.83                   \\
8-bit            & 0.65          & 97.40$\pm$ 0.24 & 88.84$\pm$0.83                     \\
4-bit            & 0.62          & 97.22$\pm$ 0.24 & \multicolumn{1}{l}{88.57$\pm$0.83} \\ \bottomrule
\end{tabular}}
\end{table}

\begin{figure}
    \centering
    \includegraphics[width=0.99\linewidth]{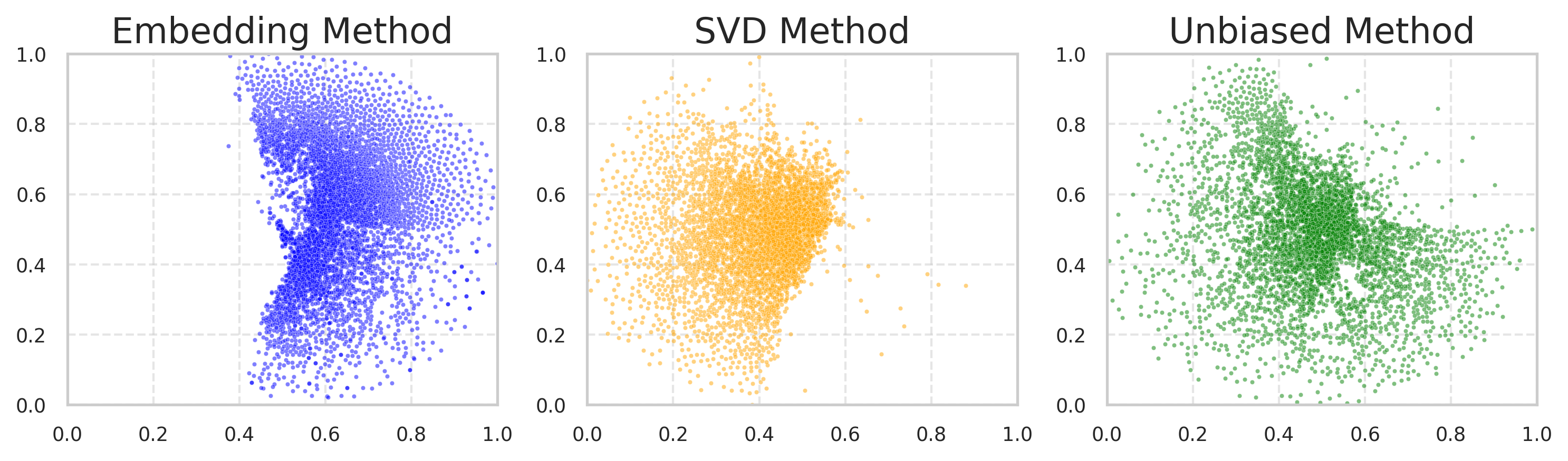}
    \caption{Visualization of sampled input embeddings from \llama-3-8B (left), the singular basis obtained through SVD decomposition of input embeddings (middle) and unbiased basis (right).}
    \label{fig:visualization_emb_basis}
\end{figure}

\bheading{Theoretical Analysis.}
We investigate the gradient magnitude for ER and TSB.
To simply notation, we denote the target \ac{is} of layer $l$, $\interstates{\llm_\cV}{l}(\inputtext))$ by $\interstatesalone$, and consider the adversary optimizes $\widehat{\bw}$ to approximate $\transformer_l(\widehat{\bw})$ to $\interstatesalone$.
Without loss of generality, we consider the squared $\cL_2$ norm for the distance $\distance$ and $\widehat{\bw}\in\Real^\inpembdim$.
The loss function is:
\begin{equation}
    \mathcal{L} =\distance(\transformer_l(\widehat{\bw}), \interstatesalone) = \norm{\transformer_l(\widehat{\bw})- \interstatesalone}_2^2.
\end{equation}

The gradient with respect to $\widehat{\bw}$ is:
\begin{equation}
\frac{\partial \mathcal{L}}{\partial \widehat{\bw}} = 2 \left( \transformer_l(\widehat{\bw}) - \interstatesalone \right)^\top \cdot \frac{\partial \transformer_l}{\partial \widehat{\bw}}.
\end{equation}
Here, the gradient magnitude of ER depends directly on the residual error $\norm{\transformer_l(\widehat{\bw}) - \interstatesalone}$ and the Jacobian $\partial \transformer_l/\partial \widehat{\bw}$ of the first $l$ Transformer layers.
For TBS, as $\widehat{\bw}=\varchangetbs(\widehat{\bz}\cdot \bB)$, the gradient with respect to $\bz$ becomes:
\begin{equation}
    \frac{\partial \mathcal{L}}{\partial \widehat{\bz}} = \frac{\partial \mathcal{L}}{\partial \widehat{\bw}} \cdot \frac{\partial \widehat{\bw}}{\partial \widehat{\bz}} = \frac{\partial \mathcal{L}}{\partial \widehat{\bw}} \cdot \frac{\partial \varchangetbs(\widehat{\bz} \cdot \bB)}{\partial \widehat{\bz}}.
\end{equation}

In the experiments, we set $\varchangetbs(\widehat{\bz} \cdot \bB)=\alpha\arctan(\widehat{\bz} \cdot \bB)$, thus the derivative is
\begin{equation}
    \frac{\partial \widehat{\bw}}{\partial \widehat{\bz}} = \frac{\alpha}{1+(\widehat{\bz}\bB)^2} \cdot\bB^\top.
\end{equation}

For small $\bz$, the scaling factor $\alpha / (1 + (\widehat{\bz}\bB)^2)\approx \alpha$, leading to larger gradient magnitudes compared to the baseline.
As $\bz$ increases, the factor $\alpha / (1 + (\widehat{\bz}\bB)^2) \to 0$, which stabilizes the gradient on $\widehat{bz}$, especially for deep model where the Jacobian can vary greatly.

In conclusion, TBS introduces a \textit{nonlinear scaling} of gradients through $\varchangetbs$, creating a dynamic where:
\begin{equation}
\norm{\frac{\partial \mathcal{L}}{\partial \widehat{\bz}}} \propto \frac{1}{1 + (\widehat{\bz}\bB)^2} \cdot \norm{\frac{\partial \mathcal{L}}{\partial \widehat{\bw}}}.
\end{equation}

\bheading{Visualization of Input Embedding and Basis.}
To support the findings of \Cref{fig:motiv-unbiased}, we also visualize the vectors through uniform t-SNE dimension reduction.
\Cref{fig:visualization_emb_basis} demonstrates that the singular basis vectors and input embeddings exhibit tight clustering in the representation space, whereas the unbiased basis displays significantly greater dispersion.

\begin{figure*}[t]
    \centering
    \includegraphics[width=0.95\linewidth]{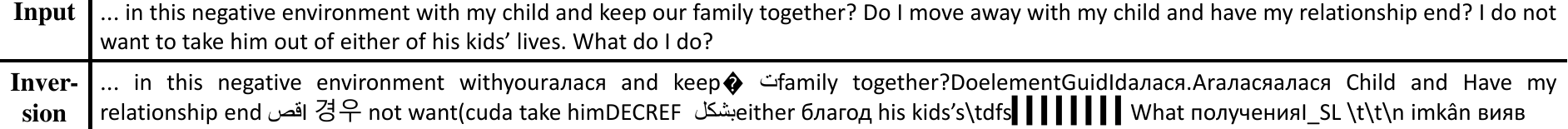}
    \caption{Failure inversion example of TBS attack in MentalHealth.}
    \label{fig:failure-tbs}
\end{figure*}

\begin{figure*}[t]
    \centering
    \includegraphics[width=0.95\linewidth]{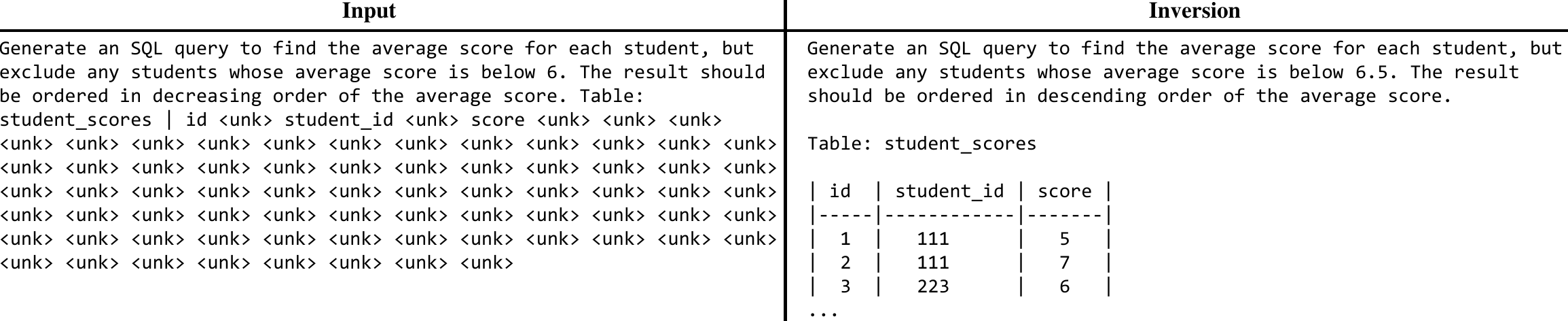}
    \caption{Failure inversion example of generation-based inversion in EvolCode.}
    \label{fig:failure-black}
\end{figure*}

\begin{figure}[t]
    \centering
    \includegraphics[width=0.95\linewidth]{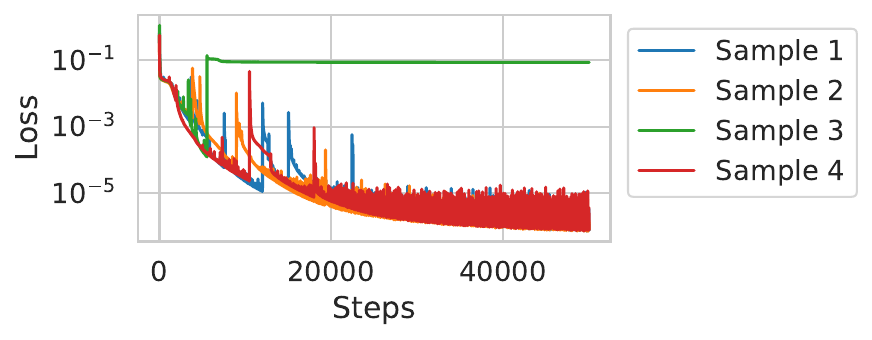}
    \caption{Loss curves of TBS attack in Mentalhealth.}
    \label{fig:failure-loss-tbs}
\end{figure}

\section{Analysis of Failure Cases}
\label{subsecapp:failurecase}

In this section, we analyze the failure cases caused by non-convergence for our optimization-based attack TBS and by the distribution mismatch for our generation-based inversion attack.

The failure cases for our white-box attack are commonly caused by inappropriate settings.
For example, too high learning rate may accelerate the optimization at the beginning but fall into local minimum afterwards, as validated by the worse inversion quality in
\Cref{fig:wb-long-lr-loss-dataset}.
A failure example of high learning rate 0.001 is shown in \Cref{fig:failure-tbs}, where we can see most inverted tokens are unreadable.
Even there are some keywords recovered, the adversary cannot guess original inputs.

Besides, due to the complexity of optimization space, we also observe failure example even under appropriate setting.
In \Cref{fig:failure-loss-tbs}, we provide the loss curves for the first four samples of Mentalhealth during TBS on \llama-3.3-70B.
We observe that the Sample 3 encountered loss divergence, leading to inversion F1 score 72.91.
We observe that Sample 3 exhibits loss divergence, resulting in the F1 score of 72.91.
In contrast, the remaining samples successfully escape local optima despite similar initial loss increases, ultimately achieving smooth convergence.
Given our fixed initialization scheme, we explore random initialization for as a potential alternative.
Empirical evaluation reveals that standard Gaussian noise initialization degrades convergence speed and ultimately yields inferior inversion performance.
Future research directions include developing bounded random initialization strategies or incorporating dynamic noise injection during optimization.

\Cref{fig:failure-black} illustrates a failure case of generation-based inversion arising from distributional mismatch.
The target text contains SQL table code that lies outside the inversion model's training distribution, resulting in uninterpretable ``<unk>'' tokens.
This demonstrates that a necessary condition for successful inversion is comprehensive training data coverage of all target input tokens.
Furthermore, the discrepancy in token distributions between the inversion model's training data and target inputs may lead to suboptimal inversion performance, particularly when significant frequency mismatches exist for critical tokens.

\begin{figure*}[t]
    \centering
    \includegraphics[width=0.99\linewidth]{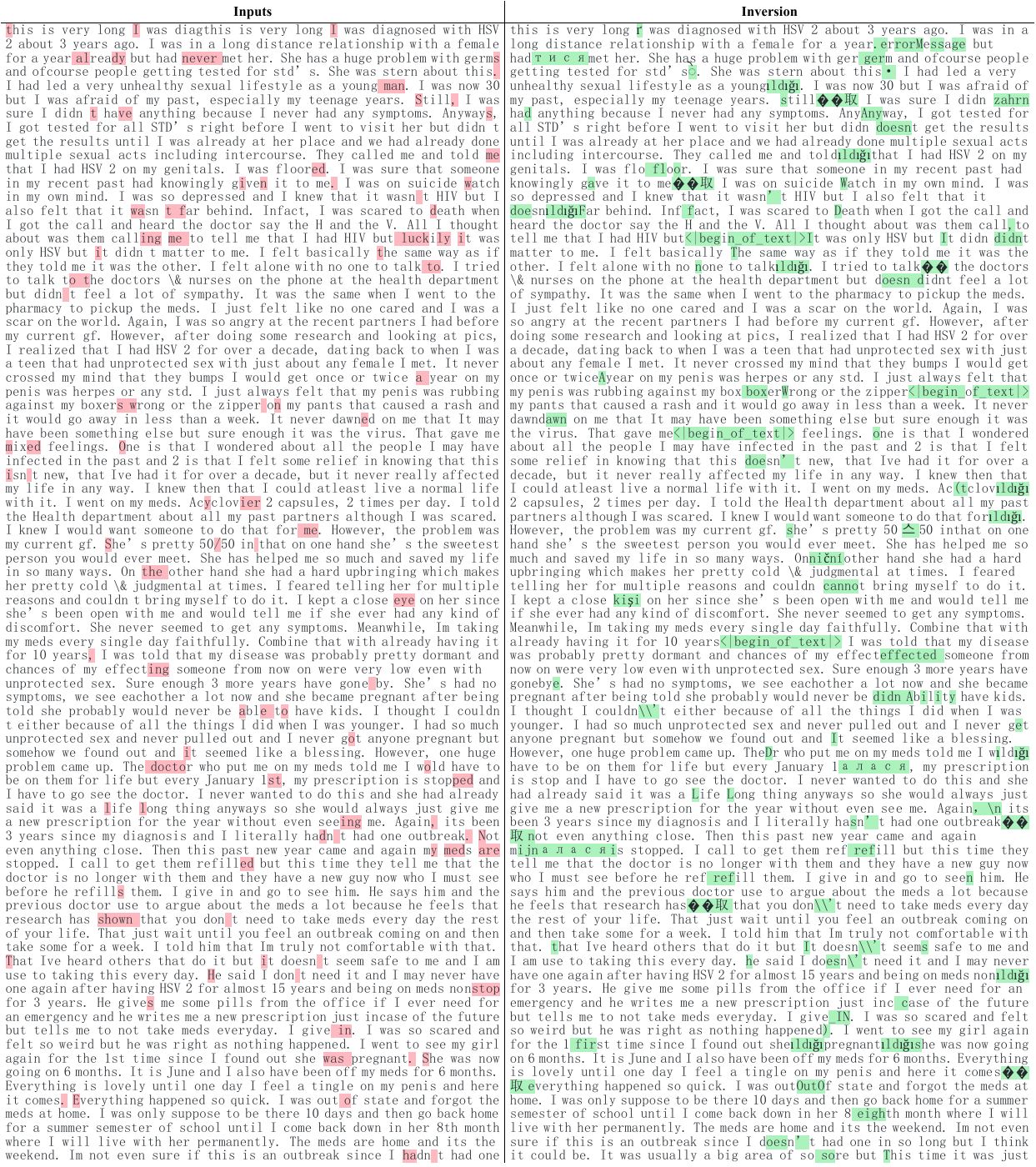}
    \caption{Inversion example of a 4,112-token prompt (Part 1).}
    \label{fig:example-p1}
\end{figure*}

\begin{figure*}[t]
    \centering
    \includegraphics[width=0.99\linewidth]{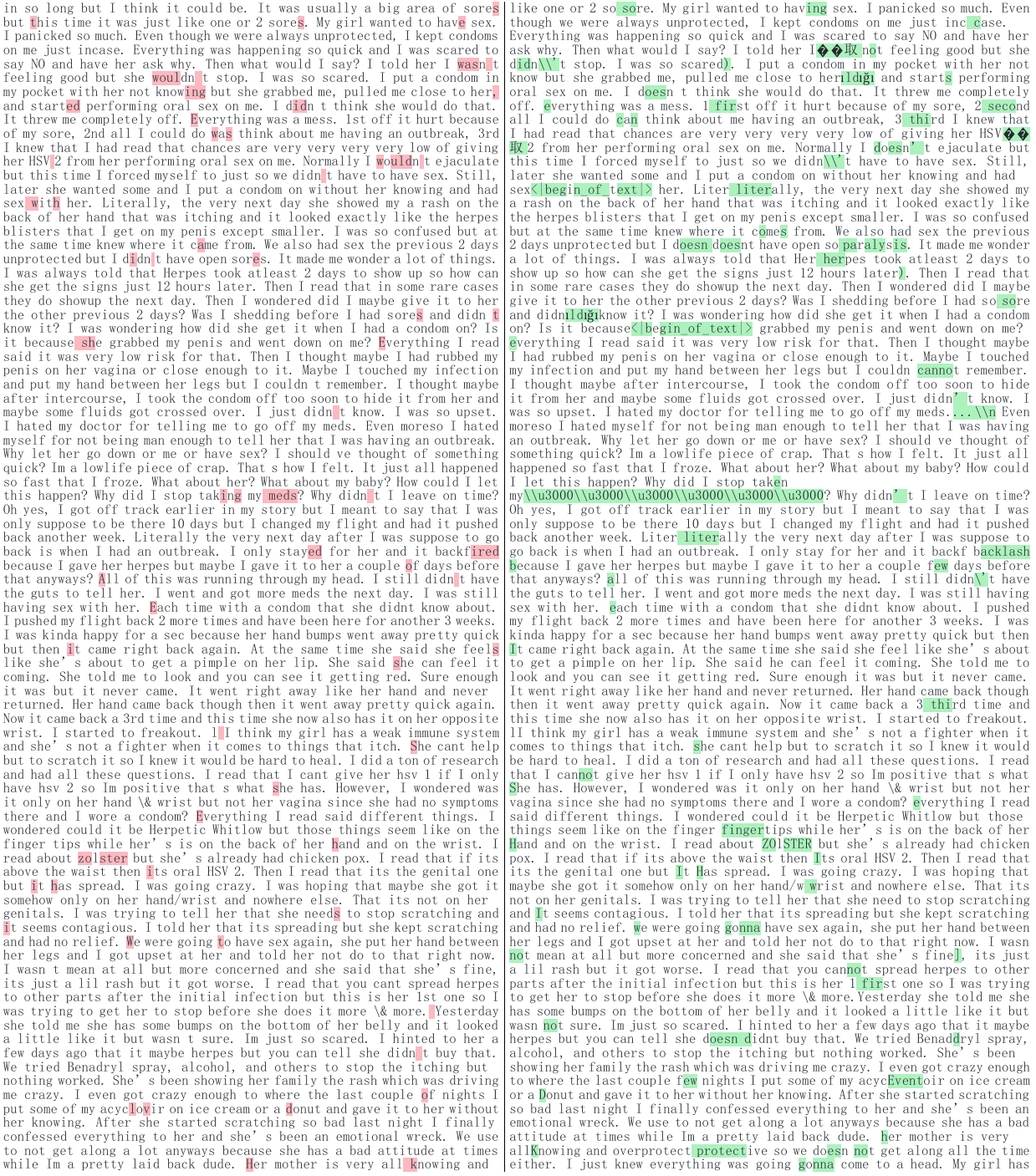}
    \caption{Inversion example of a 4,112-token prompt (Part 2).}
    \label{fig:example-p2}
\end{figure*}

\begin{figure*}[t]
    \centering
    \includegraphics[width=0.99\linewidth]{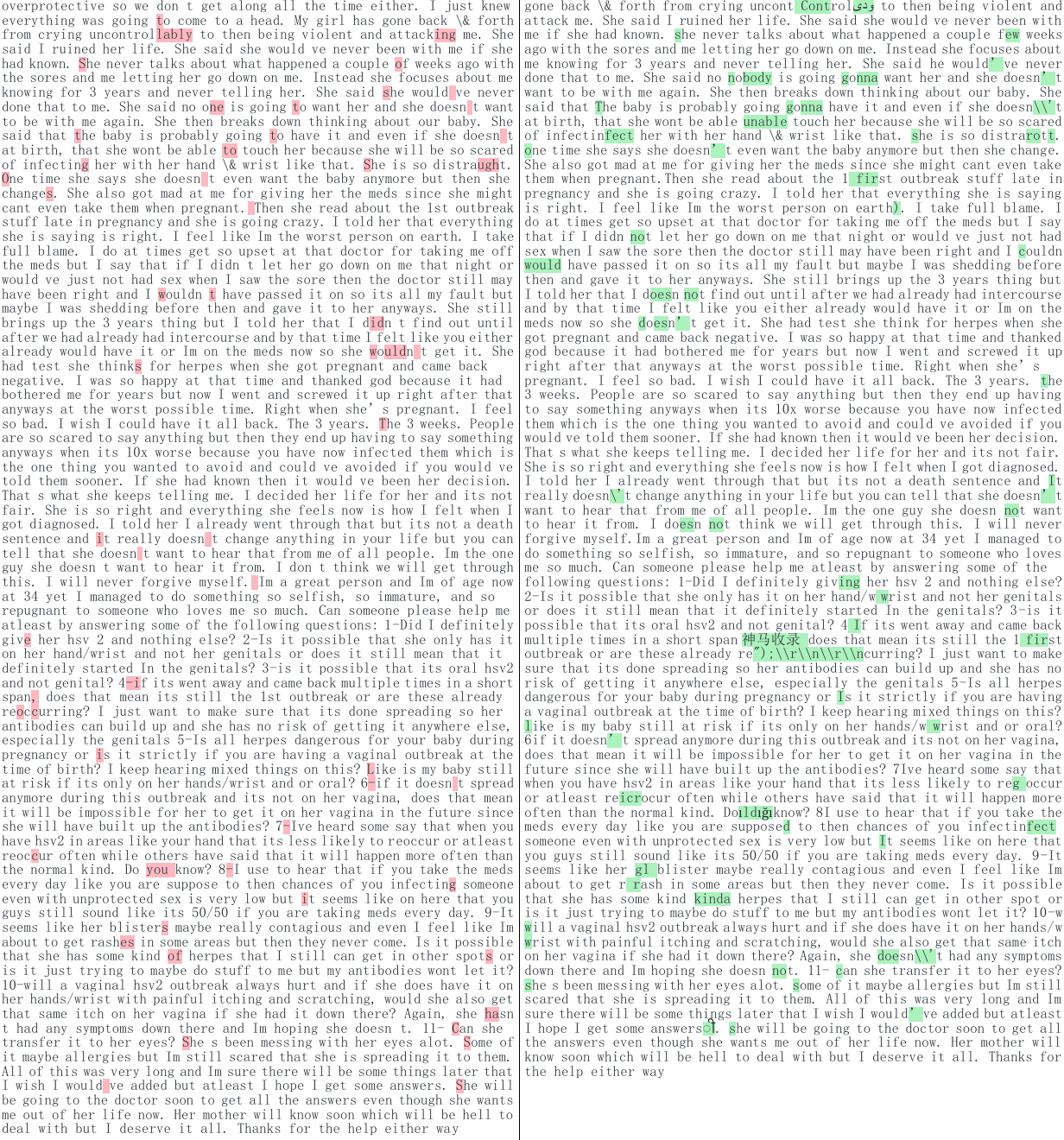}
    \caption{Inversion example of a 4,112-token prompt (Part 3).}
    \label{fig:example-p3}
\end{figure*}

\end{document}